\documentclass[twocolumn]{aastex62}

\usepackage[caption=false]{subfig}
\usepackage{natbib}
\usepackage{graphicx}
\usepackage{times}
\usepackage{amsmath,amstext}
\usepackage{appendix}
\usepackage{float}
\usepackage{rotating}
\usepackage{gensymb}

\newcommand{\sersic}{S\'{e}rsic }
\LTcapwidth=\textwidth
\usepackage{listings}
\usepackage{spverbatim}
\usepackage{fancyvrb}
\usepackage{fvextra}
\fvset{breaklines=true}

\begin{document}


\title{Limitations on Morphological Fitting for \textit{JWST} ``Little Red Dots''}

\author{Kelly E. Whalen}
\affil{NASA Goddard Space Flight Center, Code 662, Greenbelt, MD 20771,USA}

\author{Kimberly A. Weaver}
\affil{NASA Goddard Space Flight Center, Code 662, Greenbelt, MD 20771,USA}

\author{Ryan C. Hickox}
\affil{Department of Physics and Astronomy, Dartmouth College, Hanover, NH 03755, USA}

\author{Erini Lambrides}
\affil{NASA Goddard Space Flight Center, Code 662, Greenbelt, MD 20771,USA}

 \shortauthors{Whalen et al.}
 \shorttitle{LRD Morphological Fitting Limitations}

\begin{abstract}
Early results from \textit{JWST} uncover a peculiar class of objects referred to as ``little red dots'' (LRDs). The extremely compact morphology of LRDs is often invoked to point towards an AGN-dominated picture in the context of their conflicting multiwavelength properties. In this work, we assess the capability of pysersic and GALFIT---commonly used tools in LRD morphological studies---to recover input parameters for a simulated suite of LRD-like objects in the F444W band. We find that: 1) these tools have difficulty recovering input parameters for simulated images with SNR $\lesssim 25$; 2) estimated PSF fraction could be a more robust physically-motivated description of LRD compactness; and 3) almost all permutations of modeled LRDs with SNR $\lesssim 50$ cannot be differentiated from a point source, regardless of intrinsic extent. This has serious implications on how we interpret morphological results for increasingly large photometric samples of LRDs, especially at extremely high-$z$ or in relatively shallow fields. We present results of \sersic and two-component fitting to a sample of observed LRDs to compare with our mock sample fitting. We find that $\sim85\%$ of observed LRDs are PSF-dominated, consistent with the AGN-dominated interpretation. The remaining $\sim15\%$ have low estimated PSF fractions (two-component fit) and sizes $\gtrsim 150$ pc (\sersic). This morphological diversity of LRDs suggests that that the population likely is not homogeneous. It possibly has a primary subset of sources consistent with the AGN-dominated hypothesis, and a secondary population of sources more consistent with arising perhaps from extremely compact starbursts.
\end{abstract}


\submitjournal{ApJ}

\section{Introduction}
\label{sec:intro}

The first years of \textit{JWST} observations have revealed an anomalous population of high-$z$ sources that have directly challenged our understanding of galaxy and supermassive black hole (SMBH) formation and evolution. These objects, referred to as ``little red dots" (LRDs; \citealt{matt24lrd}), are ubiquitously found in \textit{JWST} blank fields, and they are characterized by extremely compact morphologies and V-shaped spectral energy distributions (SEDs) with red rest-frame optical colors and excess ultraviolet (UV) emission \citep[e.g.,][]{akin24overabundance_lrd,koko24census,labb25uncover_redAGN}. It is still largely unclear what type (or types) of astrophysical objects are primarily responsible for producing seemingly disparate multiwavelength photometric and spectroscopic signatures.

The puzzle regarding the nature of LRDs begins with trying to understand the shape of their SEDs constructed from NIRCam broadband photometry. The characteristic V-shape can be consistent with several different physical models. One scenario is that LRDs are dominated by a redenned active galactic nucleus (AGN). This model assumes that the red optical/near infrared (IR) continuum arises from emission from obsucuring material, and that the UV excess is scattered light produced by the accretion disk \citep[e.g.,][]{akin23twoLRD,koce23broadline_lrd,barr24eros,kill24lrd}. One challenge to this picture is that there is a lack of rise in the rest near-IR \citep[eg.][]{akin25uv_lines_lrd,wang25massivestarsLRDs}. An additional facet is that it seems that black holes in LRDs are over-massive compared to their host galaxies if we compare them to local scaling relations \citep[e.g.][]{rein15mbh_mstar_local, maio24jades_lrd_population}. This requires LRDs to either have evolved from extremely massive seeds or for them to have been accreting at close to or beyond the Eddington rate since they were formed \citep[e.g.][]{koko23uncover_broadline_lrd, nata24directcollapse_lrd, pacc24growth_lrd,lamb24superEdd_lrd}, which is unexpected given typical AGN duty cycles.

It is also possible that the shape of LRD NIRCam SEDs can be obtained by assuming that they are extremely compact starburst galaxies where the UV excess arises from unobscured emission from hot, young stars and the red/near-IR optical continuum is reprocessed thermal emission from dust obscured star formation \citep[e.g.,][]{pere24smiles_miri_lrd,bagg24starburst_broadline_lrd}. However, this assumption means that LRDs are extremely massive for their epoch ($M_{*} \sim 10^{10} \ M_{\odot}$), giving rise to significant tensions with $\Lambda$CDM models for cosmology \citep[e.g.,][]{boyl23highz_lcdm,deke23high_z_gal_formation}. 

NIRCam SEDs for LRDs are also consistent with composite models that attribute the UV excess to newly formed stars and the red continuum to an attenuated AGN. These models alleviate the tension of the galaxy-only scenario with $\Lambda$CDM, as not all of the luminosity has to arise from stellar emission, allowing for them to have stellar masses more typical of what is expected at this epoch, although their black holes are preferentially over-massive while compared to local relations \citep[e.g.,][]{leun24miri_sed_fits_composite_lrd,duro25overmassiveBHs}.

Spectroscopic and multiwavelength photometric observations are required to obtain better constraints on the physical natures of LRDs. Followup observations using NIRSpec reveal that up to $\sim 80\%$ of LRDs exhibit Doppler broadening of Balmer emission lines in their spectra \citep[e.g.,][]{hari23broadlineAGN_lrd,green24uncover_broadline_lrd,koko23uncover_broadline_lrd,furt24highBHtoMass,wang24broadline_outflow_lrd,hvid25_rubies-bl-lrds,koce25lrdsample}. These broad lines have typically been used to support the AGN-dominated nature for LRDs in conjunction with their apparent point-like morphology. However, it is also possible that compact galaxies with extremely high stellar densities are capable of driving the gas kinematics required to produce broad emission features \citep[e.g.,][]{bagg24starburst_broadline_lrd}. Although these seem to be exotic, we note that galaxies with extreme stellar densities on the order of that required by \citet{bagg24starburst_broadline_lrd} have been observed at $z \sim 0.5$, and are likely to be more common at high-$z$ \citep[e.g.,][]{diam21CSoutflows,whal22spacedensity}.

Under the most basic AGN unification schemes, we would expect LRDs to be unobscured enough to be X-ray detectable given that there is a line-of-sight to broad-line emitting gas \citep[e.g.,][]{urry95unifiedmodel,pado17AGN,hick18obsc-review}. However, the vast majority of LRDs remain undetected in archival, stacked X-ray imaging \citep[e.g.,][]{furt24highBHtoMass,anan24xrayStack_lrd, lamb24superEdd_lrd,yue24xray_stack_lrd,koce25lrdsample}. Given their UV luminosities as directly measured from \textit{JWST} photometry, we would expect them to be powerful X-ray sources under the assumption of accretion by an optically-thick, geometrically-thin distribution \citep[e.g.][]{luss16xrayUV,biso21quasarUVxray,sign23UVxray,yue24xray_stack_lrd,maio25chandra_blr_lrd,anan24xrayStack_lrd}. Within the AGN framework, it is possible that LRDs are growing at super-Eddington accretion rates, thus having their X-ray emission intrinsically suppressed \citep[e.g.,][]{lamb24superEdd_lrd, pacc24super_eddington_lrd,lupi24superEddhiz}. It has also been proposed that dense, broad line emitting gas near the AGN could be responsible for providing Compton thick column densities that could absorb X-ray emission \citep[e.g.][]{maio25chandra_blr_lrd}. Recent works have also suggested that LRDs could be accreting SMBHs residing in dense envelopes of Compton thick ionized gas \citep[e.g.,][]{degr25black_hole_star,naid25black_hole_star, rusa25black_hole_star}. Though, recent results showing either high-ionization or rare Fe transitions challenge a simple high-covering fraction, dense gas envelope interpretation \citep[e.g.][]{lamb25LRDfe}.

The lack of X-ray emission could also be consistent to LRDs being dominated by a compact starburst, as they would be too intrinsically X-ray weak to be detectable from high-$z$ in deep, archival X-ray image stacks.
The rest frame infrared (IR) potentially has the power to constrain the nature of LRDs, as we would expect the IR slope to flatten out in the near-IR for dusty star forming galaxies, but for it to continue to rise for AGN due to hotter dust temperatures in the obscuring torus compared to that in the ISM for galaxies. A subsample of LRDs have been observed using the MIRI instrument on \textit{JWST}, and their photometry reveal that their SEDs in fact flatten out at around rest-frame 1 $\mu$m \citep[e.g.,][]{pere24smiles_miri_lrd,will24extremelyRedGalaxies_lrd,sett25nodustLRD}. This is largely more consistent with LRDs being dominated by stellar emission, although composite models including dust-free AGN cannot be completely ruled out \citep[e.g.,][]{leun24miri_sed_fits_composite_lrd}. Additionally, sub-millimeter observations reveal that LRDs do not contain detectable amounts of cold dust \citep[e.g.,][]{case24alma_lrds,case25almanodetect60lrds,sett25nodustLRD}. Assuming that their red optical emission arises from starlight is in conflict with most dust formation models, suggesting that optical emission from LRDs likely has non-stellar origins, in tension with the MIRI data.

In the context of LRDs exhibiting conflicting signatures of AGN and star formation dominance in their multiwavelength data, their compact morphologies have often been used to invoke their likely AGN nature on the basis that the galaxies consistent with these sizes would have too high of a stellar mass density. As mentioned earlier, including some AGN emission in addition to galaxy light in LRD models would alleviate some of the pressure to be so massive. Additionally, assuming a top-heavy initial mass function (IMF) could also allow for LRDs to have lower derived stellar masses while maintaining their brightness \citep[e.g.,][]{wood24topheavy_imf,wang24highz_imf}.

In light of all of their peculiar observed multiwavelength properties, a coherent model for the full population of LRDs is highly contested. Deeper inspection of some of the assumptions on what defines an LRD is needed. In this work, we inspect the robustness of the morphological LRD criteria. Use of parametric fitting codes to quantitatively describe LRD morphologies have given effective radius ($R_{\textrm{eff}}$) measurements that range from as small as $\sim 30$ pc to as large as $< 200$ pc within F444W images \citep[e.g.,][]{koko24census,koce25lrdsample}. This is a fairly large dynamic range, with objects on the smaller end being more compatible with the AGN-dominated interpretation for the nature of LRDs and those on the larger end being consistent with the sizes of massive compact starbursts \citep[e.g.,][]{sell14CSoutflow}. The extreme compactness of these objects certainly push against the limitations of the computational tools that are used to measure morphology, as their sizes are on the order of the width of the instrumental point spread function (PSF). It is imperative that we fully understand the limitations of the tools we use before we can interpret the estimated sizes of LRDs.

In this work, we aim to address several questions about the limitations of commonly-used morphological fitting routines with regards to determining the sizes of LRDs:
\begin{enumerate}
    \item {What are the size and signal-to-noise  ratio (SNR) limits at which morphology fitting codes can no longer recover the intrinsic effective radius, assuming that the observed source is represented as a \sersic profile?}
    \item{Assuming that a LRD can be represented as a point source embedded in a more extended galaxy (consistent with AGN + compact starburst scenario), can morphological fitting codes be used to accurately perform bulge/disk decomposition?}
    \item{Are LRDs distinguishable from point sources? If so, under what set of model assumptions?}
\end{enumerate}

We note that LRD is a term that has been used to describe a potentially disparate sample of objects in the literature that have been selected using various criteria. For the sake of this work, we use the term LRD to refer to targets that are photometrically selected based on their rest UV and optical colors that have also fulfilled a compactness cut ($F_{0\farcs04}/F_{0\farcs02} < 1.7)$. We do not require them to have broad lines.

The outline of the paper is as follows: in Section 2 we discuss the construction of mock LRD F444W cutouts. In Section 3 we discuss the morphological fitting codes we are testing and the results to running them on our suite of simulated images. In Section 4 we contextualize the results of this fitting by comparing the limits we derive to observed samples of LRDs . We adopt a cosmology of $H_{0} = 70.2 \ \textrm{km} \textrm{s}^{-1} \textrm{Mpc}^{-1}$,
$\Omega _{M} = \Omega _{CDM} + \Omega _{b} = 0.229+0.046 = 0.275$, and
$\Omega _{\Lambda} = 0.725$ \citep{koma11WMAP}


\section{Modeling LRD cutouts}
In this section, we describe in detail how we construct mock image cutouts of LRDs that are representative as those found in observed \textit{JWST} fields. The DAWN \textit{JWST} Archive\footnote{https://dawn-cph.github.io/dja/index.html} (DJA) is a publicly available repository for fully reduced \textit{JWST} data, including deep field mosaics. Data published on the DJA has been widely used in high-$z$ AGN and galaxy studies, including the census of LRDs presented in \citet{koko24census}, which are adopting as the parent sample in this work. We base our simulated LRDs on cutouts from the DJA mosaics, and we convolve all of our modeled LRDs with the empirical PSFs available for download on the DJA repository.

We model the LRDs themselves as the linear combination of an extended, \sersic component with varying effective radius ($R_{\textrm{eff}}$) and \sersic index ($n$), and a PSF component, where the relative contribution to the surface brightness profile by each component is set by a parameter we refer to as $f_{\textrm{PSF}}$. We convolve each modeled LRD with a DJA empirical PSF that will also be passed as the PSF model in the fitting routine. We then add noise to the images to obtain a desired signal-to-noise ratio (SNR). We generate cutouts for permutations of LRDs with model parameters within the ranges: $0.25 < R_{\textrm{eff}}/\text{pixel} < 2$, $1 < n_{\textrm{sers}} < 4$, $0 < f_{\textrm{PSF}}< 1$, and $10 < \text{SNR} < 250$.

\subsection{Extended stellar component}
We begin modeling our the LRD cutouts using the tools available in the Python package, GalSim \citep{rowe15galsim}. GalSim is a flexible, open-source package that allows for the simulation of astronomical images for a variety of ground and space-based instruments. Galfit has a variety of base surface brightness models available that are binned for each pixel in the simulated image. We model each simulated galaxy's extended stellar component as a \sersic profile whose brightness as a function of radius from the center ($R$) is given as:
\begin{equation}
    I_{\text{\sersic}}(r) = I_{e} \exp \bigg\{ -b_{n} \bigg[ \bigg(\frac{R}{R_{\textrm{eff}}} \bigg)^{1/n} - 1 \bigg] \bigg\}.
\end{equation}
where $n$ is the \sersic index, or steepness, of the profile, $R_{\textrm{eff}}$ is the effective radius that contains half of the integrated light of the galaxy, $I_{e}$ is the brightness at $R_{\textrm{eff}}$, and $b_{n} \sim 2n - 1/3$ for $n < 8$. We choose GalSim over Astropy's \citep{astropy} Sersic2D package because we are working with very small targets, and Sersic2D does not oversample at its center which leads to a significant discrepancy between the peak flux and the integrated flux in the central pixels for small, high-$n$, profiles\footnote{https://github.com/astropy/astropy/issues/11179}. For each of our simulated stellar components, we vary $R_{\textrm{eff}}$ and $n$, and fix $I_e$ such that the integrated flux for the entire profile is one, and set the inclination and ellipticity to zero. We then convolve the simulated images of our \sersic profiles with the \textit{JWST} F444W ePSF calculated in the previous subsection.  

\subsection{Central point source}
In this work, we consider the potential composite nature of LRDs where there is a central point source embedded within a more extended stellar distribution as described in the previous subsection. We account for this by treating the total surface brightness profile as a weighted sum of the extended stellar distribution and a point source component modeled by the PSF profile in \textit{JWST} F444W. This is given as 
\begin{equation}
    I_{\textrm{LRD}} (R) = f_{\textrm{PSF}} \cdot I_{\textrm{PSF}}(R) + (1-f_{\textrm{PSF}}) \cdot I_{\text{\sersic}}(r).
\end{equation}
We vary $f_{\textrm{PSF}}$ while constructing our simulated LRDs to determine how the effectiveness of morphological fitting routines changes as the central point source begins to dominate the surface brightness profile. We then renormalize $I_{\textrm{LRD}}(R)$ such that the integrated flux within a 0".36 diameter aperture is equal to 1 $e^{-}$, allowing us to easily rescale it later when we inject noise to obtain a specified signal-to-noise ratio (SNR).


\subsection{Noise}
\label{subsubsec:noise}
Once we construct baseline surface brightness profile images for our simulated LRDs, we then inject noise. We consider contributions from read nose, dark current, source Poisson noise, and background noise while constructing our cutouts. We treat the read noise and the dark current noise as a single effective noise component that is modeled as a randomly generated image where the value in each pixel is drawn from a Gaussian distribution with a mean $\mu = 0$ e$^{-}$, and standard deviation $\sigma = 9.28$ e$^{-}$, as specified in the NIRCAM performance documentation\footnote{https://jwst-docs.stsci.edu/jwst-near-infrared-camera/nircam-instrumentation/nircam-detector-overview/nircam-detector-performance\#gsc.tab=0}. The dark  noise component assumes a kilosecond exposure, so we scale it to match the exposure time of the DJA CEERs mosaic ($\sim9.5\times10^{4}$ seconds). We convert the effective noise to units of analog-to-digital units (ADU) using the gain (1.82 e$^{-1}$/ADU). The Poisson noise from the source is similarly a randomly generated image, but instead drawn from a Poisson distribution where the noise in each pixel is $\sim \sqrt{N_{counts}}$. 

The most dominant source of noise in our cutouts is from the background. Using the \textit{JWST} background tool, we computed an average background flux for the central coordinates of the CEERS mosiac ($214.92\degree$, $52.87\degree$) to obtain background levels representative of that in the observed images of LRDs. Depending on the time of year for this particular mosaic, the background flux levels range from 0.24-0.34 MJy/steradian. For simplicity, we assume a that the background is uniform across the image at its mean value of 0.29 MJy/steradian. We convert the background flux to counts/s using conversion factor \textsc{photmjsr}, and then to counts by multiplying it by the effective exposure time for the mosaic. We then calculate the background noise image by randomly drawing from a Poisson distribution at each pixel. Our analysis requires that our cutouts be background subtracted, so we then subtract the mean background value out from the background noise image. 

We make note of the fact that we are trying to simulate cutouts from a mosaic that has been reprocessed and ``drizzled'' \citep{fruc02drizzle}. The Drizzle algorithm linearly reconstructs under-sampled images to increase their spatial resolution. For example, NIRCAM F444W images natively have a scale of 0.063"/pixel, but the processed mosaics have been reconstructed to have that of 0.04"/pixel. The algorithm works by shrinking then mapping input pixels onto a sub-sampled output grid. This has the effect of introducing correlated noise, since the noise associated with one input pixel now gets spread out over several sub-sampled pixels. This means that the signal-to-noise ratios of sources in images would be over-estimated without including a correlated noise correction \citep[e.g.,][]{bagl24uvlf_corrnoise}. 

To ensure that our simulated galaxy cutouts are representative of those constructed from observed mosaics, we must introduce correlated noise into the simulated noise images we described above. We do this by making cutouts of several LRD candidates included in \citet{koko24census} from DJA. For each cutout, we mask the LRD and contaminant sources, and then compute the autocorrelation function (ACF) of the background. For each simulated LRD, we convolve the background noise image with the ACF measured from actual background cutouts to redistribute the power of the simulated background noise to make it more representative of what we would expect from drizzled data with a correlated noise structure.

We are fundamentally interested in determining if there is a SNR limit at which morphological fitting becomes unreliable for extremely compact sources, such as LRDs. In order to address this, we need to rescale the modeled LRD surface brightness profiles and add them to the simulated correlated noise images. The SNR for any object measured via aperture photometry is given as:
\begin{equation}
    SNR = \frac{F_{aper}}{\sigma_{aper}}
\end{equation}
where $F_{aper}$ is the measured flux in a 0".36 diameter aperture and $\sigma_{aper,tot}$ is the total noise. The noise can be further broken up into
\begin{equation}
    \sigma_{aper,tot} = \sqrt{F_{aper} + \sigma_{sky,corr}^{2}} 
    \label{eq:tot_noise_uncorr}
\end{equation}
where the first term is the source Poisson noise contribution, $\sigma_{sky,corr}$ is the sky noise corrected for correlation. This accounts for the background Poisson noise and the baked-in CCD noise such as read noise and dark current noise. However, when aperture photometry is performed on an image, the measured noise has not yet been corrected. We must first compute $R$, a correction factor that is used to scale the measured aperture noise to account for correlation between pixels. This allows us to rewrite Equation \ref{eq:tot_noise_uncorr} as 
\begin{equation}
\sigma_{aper,tot} = \sqrt{F_{aper} + R \ \sigma_{sky,uncorr}^{2}} \ .
\end{equation}

In an image with completely uncorrelated pixels, the measured sky noise within an aperture will scale as a factor of $\sqrt{N_{pixels}}$. In an image where all of the pixels are perfectly correlated, the sky noise within an aperture scales as a factor of $N_{pixels}$ \citep[e.g.,][]{quad07corrNoise}. For situations like our own where the noise is partially correlated, the noise will scale as 
\begin{equation}
\sigma_{N} \sim \sigma_{b} ( N_{pixels}^{\beta})
\label{eq:beta}
\end{equation}
where $0.5 < \beta < 1$, $\sigma_{N}$ is the measured aperture sky noise and $\sigma _{b}$ is the standard deviation of the pixels only containing sky background. We can compute $\beta$ following the methodology described in \citet{papo16spitzerSurvey_corrNoise} (see also \citealt{labb03HDFsouth_noise,blan08MUSYC_noise,whit11z3bimod}). In short, we can estimate $\beta$, by placing 1000 non-overlapping, random apertures with radii $0.5 < r < 12$ pixels in the regions of DJA mosaics that are free of sources and/or bad pixels. In each aperture, we measure the fluxes to compute the normalized median absolute deviation, which can be a proxy for the measured noise within an aperture of a given size ($\sigma_{N}$) \citep[e.g.,][]{beer90absMedFluxdev}. We then linearly fit Equation \ref{eq:beta} to obtain an estimate of $\beta$. 

The uncertainty values given while performing aperture photometry assume that the images are perfectly uncorrelated, or have $\beta = 0.5$. Fitting Equation \ref{eq:beta} gives the appropriate scaling for aperture noise as a function of aperture size for data with a given amount of pixel correlation. Since $R$ is the correction factor between the assumed to be uncorrelated noise values and the real, correlated aperture noise, it can be expressed as $R = N_{pixels}^{\beta - 0.5}$. Therefore, the SNR as calculated by aperture photometry is given as 
\begin{equation}
    SNR = \frac{F_{aper}}{\sqrt{F_{aper} + R \ \sigma^{2}_{sku,uncorr}}}.
\end{equation}
We can solve the following quadratic to obtain an expression for the aperture flux of an object as a function of SNR:
\begin{equation}
    F_{aper} = \frac{SNR^{2}}{2} \bigg(1 + \sqrt{\frac{1 + 4 \ R\ \sigma_{sky,uncorr ^{2}}}{SNR^{2}}}\bigg)
\end{equation}
Using this expression as well as the correlated noise maps we constructed, we scale the surface brightness profile image to have the aperture flux to needed to achieve the desired SNR for a particular modeled LRD. We then recompute the source Poisson noise to account for the rescaled flux and add that back into the noise image. We note that the actual SNRs of these simulated LRDs are a $\sim 2$\% lower than the desired SNR due to the rescaling of the source Poisson noise.

The morphological fitting codes require us to pass an RMS image along with each object cutout. We simulate the root-mean-square (RMS) image for each model LRD by generating 500 realizations of a possible noise image given all of the LRD's input parameters. We then take the root mean square of these realizations and assign the resulting image as the RMS image. We also note that the per-pixel noise in the RMS image is likely a lower-limit on the noise, since it does not explicitly account for the pixel-to-pixel correlation introduced by Drizzle. We adopt the CEERS correlation factor of $R=2.86$ for our mock objects.

\section{Fitting the simulated LRDs}
\begin{figure*}
    \centering
    \includegraphics[width=0.9\linewidth]{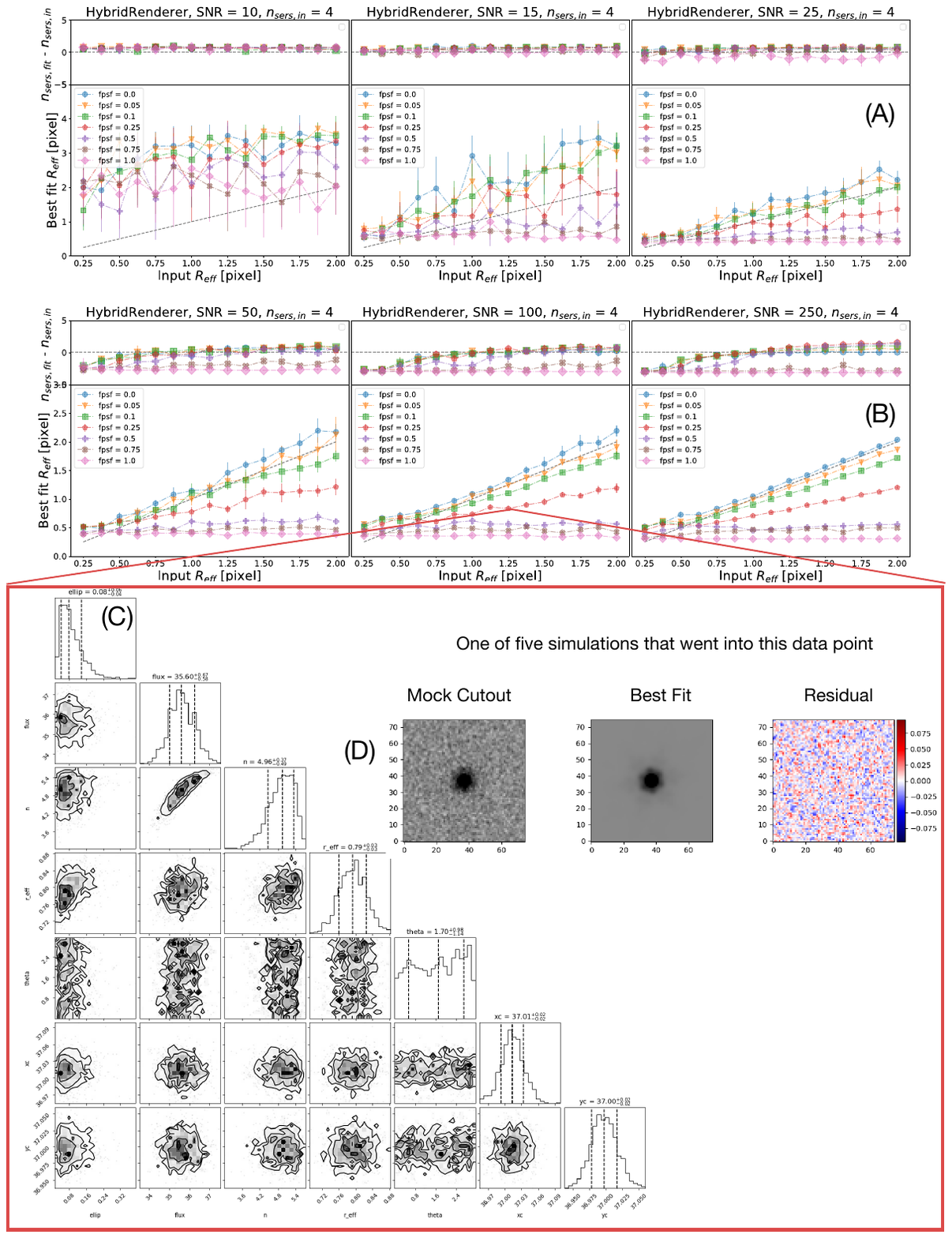}
    \caption{\sersic fits to $\text{SNR} \lesssim 25$ LRDs do not yeild reliable $R_{\textrm{eff,fit}}$ estimates. Results from pysersic single-component \sersic fits to our sample of mock LRDs at varying $\text{SNR}$. Each of the points in \textit{Panels (A)} and \textit{Panels (B)} represents a particular mock LRD image that is being fit. All one-to-one relationships between input and best-fit parameters are shown as grey, dashed lines. We note that we only expect fits to the $f_{\textrm{PSF}}=0$ images to match these. \textit{Panels (A)}: The top of each of the six panels presents residuals between the best-fit $n_{\textrm{sers}}$ and the input used to generate a given mock image. \textit{Panels (B)}: Best-fit values of $R_{\textrm{eff}}$ versus the input. Along with \textit{Panels (A)}, uncertainties are large and the best-fit values do not begin to converge towards the input values until $\text{SNR} \gtrsim 50$. \textit{Panel (C)}: Example corner plot for one of the $\text{SNR}=100$ images that was well-fit by the \sersic model. \textit{Panel (D)}: The residual image computed with the mock LRD image and the best-fit model.  }
    \label{fig:pysers_hyb_sers_n4}
\end{figure*}

The overarching aim of this work is to determine the limitations of well-known morphological fitting codes in addressing several questions regarding the nature of compact targets including JWST LRDs. Answering each of these questions requires a slightly different approach. 

The first question addressed in this work, as mentioned in Section \ref{sec:intro}, aims to determine the minimum size for a galaxy that a particular morphological fitting code can accurately recover the input parameters that were used to create that given mock cutout. This is a direct comparison to the LRD size estimates that have been presented in the literature. We approach this objective by conducting a basic \sersic fit for each of the cutouts we have generated.

The second question aims to determine for what subset of input parameters that popular morphological fitting codes could resolve a central point source from a small, extended, stellar component. One major challenge with understanding LRDs is that it is difficult to determine whether their emission is dominated by an AGN, a compact galaxy, or that it has some significant contribution from both. Being able to separate a LRD into individual components would allow us to determine if the underlying stellar population is actually extremely compact or if it is a more ``normal'' galaxy that hosts a bright AGN. Additionally, obtaining an estimate on the fraction of light contained in an unresolved component would allow for a much more robust study of the stellar populations residing in the extended component, as it would provide tighter constraints on an ``AGN fraction'' parameter in SED fitting. We address this question by performing a similar morphological fit, but with a secondary PSF component included in the model used for fitting.

Lastly, we aim to determine if there is a set of model parameters for which which our mock LRDs are indistinguishable from point sources. JWST imaging reveals that LRDs are largely unresolved. Compact \sersic profiles that have been convolved with the PSF are most easily distinguished from a true point source in the wings of the profile; which makes differentiating between the two difficult since the wings could be too faint to detect over noise in the image. We perform PSF fits to our mock LRDs to assess whether or not they are distinguishable from a pure PSF model.

In the following subsections, we describe each of the fitting codes we include in this work, as well as the priors we use for each of the models we fit to our mock LRD observations.

\subsection{pysersic}
One of the primary codes we used to conduct this analysis was pysersic \citep{pash23pysersic}. Pysersic is a Bayesian framework parametric fitting code that implements a Markov Chain Monte Carlo (MCMC; e.g., \citealt{metr53metropolishastings}) routine to robustly sample the posterior distributions for each of the parameters that make up a given morphological model. Pysersic includes several built in surface brightness models that can be fit to actual cutouts. In our analysis, we use the \sersic, \sersic plus point source, and the pure point source. The Python package is built using jax \citep{brad18jax}, and implements numpyro for its Bayesian inference \citep{numpyro_1,numpyro2}. Pysersic was utilized to obtain the LRD size upper limits presented in \citet{koko24census}, which is one of the only works in the literature that has quantitatively reported limits on LRD sizes.

For consistency with the size upper limit analysis in \citet{koko24census}, we assume uniform priors on effective radius ($ 0.25 < R_{\textrm{eff}} <5$ pixels) and \sersic index ($0.65 < n < 6$). We found that pysersic experienced difficulties trying to find the centroid for low SNR sources. To combat this, we additionally included a uniform prior on $xcen$ and $ycen$ between 36 and 39 pixels. We estimate parameter values and uncertainties using the "svi-mvn" method. This is different from the Laplace approximation used in \cite{koko24census}, but we found that there is not a significant difference between the results obtained from each method. We chose this because the ``svi-mvn" method samples the posterior distributions for the parameters being fit, allowing us to determine if there are degeneracies between any of the model parameters. We also note that pysersic can construct best fit models using either a ``PixelRenderer" that renders the image in pixel space and then convolves it with the appropriate PSF, or a ``HybridRenderer" that utilizes the approach described in \citet{lang20hybridrenderer} to approximate the radial profiles and PSFs in Fourier space as series of Gaussians. The `HybridRenderer" returns the resulting image in pixel space using an inverse fast-Fourier transformation. We carry out all our pysersic fits using both the Pixel and Hybrid renderers. We note that there is not a significant difference between their quality of fits.

\subsection{GALFIT}
GALFIT \citep{peng02galfit, peng10galfit2} is a 2-D parametric galaxy morphology fitting code. It is highly flexible; it offers a variety of parametric models that can be fit to the data and allows for multi-component fits. GALFIT is a non-linear least-squares fitting algorithm written in C that uses the Levenberg-Marquardt technique to minimize residuals on a given fit. It has been widely utilized in galaxy morphological fitting since its inception \citep[e.g.,][]{shet10spitzer-galfit,meer15sdss-galfit,schu17bh-bulge-galfit}, so we incorporate it in this work to compare its performance in fitting the morphology of LRDs to that of pysersic. For all of the fits, we assume uniform priors on effective radius ($0.25 < R_{\textrm{eff}}< 5$ px) and  \sersic index ($0.65 < n < 6$). Every other model parameter is left free in the fitting.

\subsection{Fit Results}
We present results from using popular morpholigical fitting codes, pysersic and GALFIT, to fit a suite of mock LRDs to determine their accuracy in estimating sizes for extremely compact galaxies. Overall, we find that pysersic and galfit behave similarly, so we will present results for the pysersic ``HybridRenderer'' fits here, and include the analogous figures for the ``PixelRenderer" and the GALFIT fits in an Appendix.

\subsubsection{\sersic fits}

First, we present results obtained by fitting our sample of mock LRDs with a single-component \sersic model. For brevity, Figure \ref{fig:pysers_hyb_sers_n4}  shows the ``HybridRenderer'' fits for all of the $n_{\textrm{sers,in}}=4$ LRDs in the sample.  Each subplot contains fits to all of the mock LRDs of a given SNR. The tops of these subplots, denoted as (A), gives the residual between the best-fit and input values of $n_{\textrm{sers}}$ for each given image. The bottom panels (B) compare the best-fit and input $R_{\textrm{eff}}$ values. The grey, dashed lines in (A) and (B) denote the one-to-one relationship between the input and best-fit values. Each point in these plots represents the fitting of a particular mock LRD image in our sample. The fits are color-coded by their input $f_{\textrm{PSF}}$ values. We note that since this a single-component \sersic fit, that we only expect the fits to the $f_{\textrm{PSF}}=0$ images to fall along the one-to-one lines. For the pysersic fits, the ``svi-mvn'' method allows us to sample the posterior while fitting (Figure \ref{fig:pysers_hyb_sers_n4} (C)). We also computed residual images for each of the mock cutouts; an example can be seen in \ref{fig:pysers_hyb_sers_n4} (D)). 

Generally, the ``HybridRenderer'' and ``PixelRenderer'' in pysersic yield comparable results, so we will discuss them together. The main takeaway from the \sersic profile fitting to the mock cutouts is that pysersic finds significantly more success fitting higher SNR cutouts. In the lowest SNR bin, almost any best fit model upon which pysersic converges is generally consistent within the uncertainties dictated by the simulated RMS image. This means that for $\textrm{SNR} \sim 10$ sources, the best fit model for a galaxy with $R_{\textrm{eff,in}} \sim 2$ pixel is nearly indistinguishable from that of a pure point source. As SNR increases, the best \sersic fits for simulated galaxies with different intrinsic surface brightness profiles become increasingly discernible from one another. However, within the $15 \lesssim \textrm{SNR} \lesssim 25$ range, there is still significant overlap between the best fits to extended targets with modest PSF contribution  and those that are more intrinsically compact. Within this SNR range, extra care must be taken while interpreting the best fit sizes measured with pysersic.

\begin{figure*}
    \centering
    \includegraphics[width=0.8\linewidth]{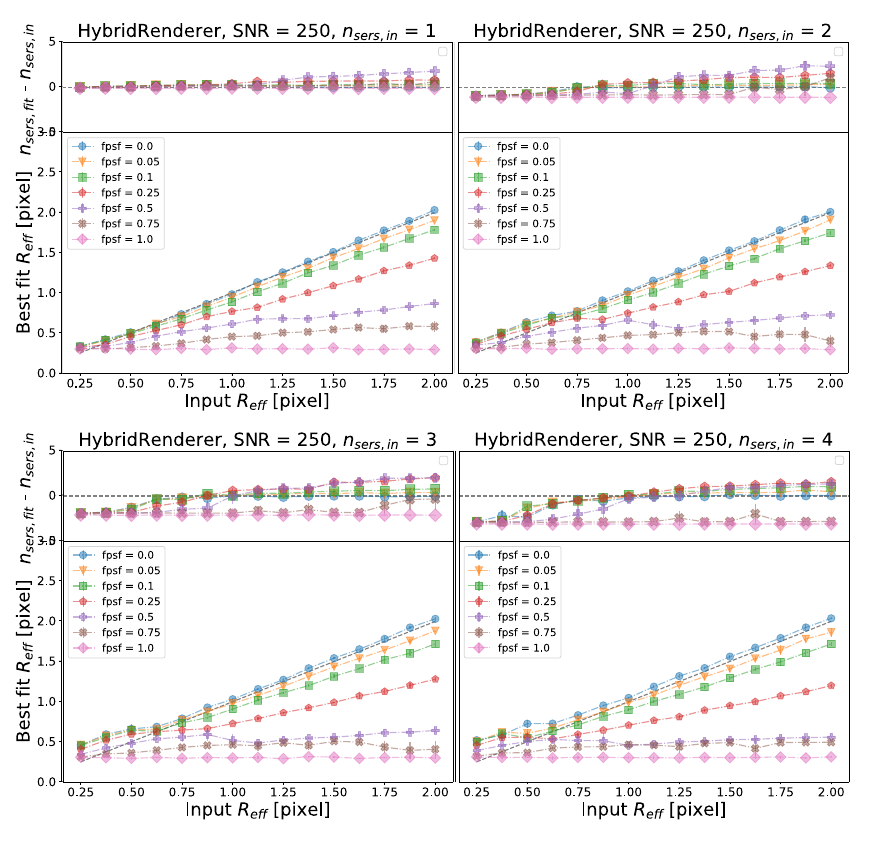}
    \caption{Pysersic has difficulty fitting simulated LRDs with steeper (larger-$n_{\textrm{sers}}$) profiles and with small effective radius. Here we present esults from single-component \sersic fitting with pysersic to $\textrm{SNR}=250$ sources with intrinsically different underlying \sersic profile shapes.  This effect propagates to the results from fitting lower-SNR simulated LRDs as well, as seen in the appendix.}
    \label{fig:nsers_fitcheck}
\end{figure*}

We also tested if there is a dependency between goodness of \sersic fit performed by pysersic and the intrinsic shape ($n_{\textrm{sers}}$) of the input simulated source. The results of this are given in Figure \ref{fig:nsers_fitcheck}. We show the suites of fits to the highest SNR simulations so any possible deviations between the input and output parameters would clearly be due to the spatial resolution limitations to pysersic rather than the extended features being lost in the noise. We find that pysersic has an more reliably recovers the true parameter values for simulated targets whose \sersic components are modeled with a less-steep $n_{\textrm{sers}}$, especially for input models consisting of \sersic profiles with small $R_{\textrm{eff}}$. This can mainly be seen in with the blue points in each of the panels, as these points represent models that consist of only a \sersic component; we would expect pysersic to be able to perfectly recover their input parameter values. However, we see this is only true for the $n_{\textrm{sers,in}}=1$ models, as there is no residual in the $n_{\textrm{sers}}$ fits across the entire $R_{\textrm{eff,in}}$ range, nor is there deviation from the one-to-one $R_{\textrm{eff}}$ output/input relation. For the other three panels where $n_{sers, in} > 1$, we find that pysersic prefers fitting \sersic profiles where $n_{sers, fit} < n_{\textrm{sers,in}}$ for models with small $R_{\textrm{eff,in}}$. This has the secondary result of pysersic consistently overestimating the sizes of input profiles with $R_{\textrm{eff,in}} < 0.65$ pixels. This trend is also seen in fits using pysersic's PixelRenderer.

As mentioned above, we additionally fit all of our mock LRD cutouts with GALFIT. The results of this can be seen in the Appendix. In short, we see the same trends as we do with our pysersic fitting. Among the major differences between the two methods is that accuracy of GALFIT's best fits does not degrade as strongly with increasing $n_{\textrm{sers,in}}$; it does not overestimate the sizes of sources with intrinsically small $R_{\textrm{eff,in}}$ even at high-SNR. We also find that the fit uncertainties are lower at $\textrm{SNR} \lesssim 50$ for pysersic, but that GALFIT has smaller uncertainties for higher-SNR cutouts. Our recommendation is the same for GALFIT as it is pysersic: care must be taken when interpreting output $R_{\textrm{eff}}$ measurements, especially for images with modest SNR.

\subsubsection{Two-component decomposition}
\label{ssec:decomp}
\begin{figure*}
    \centering
    \includegraphics[width=1\linewidth]{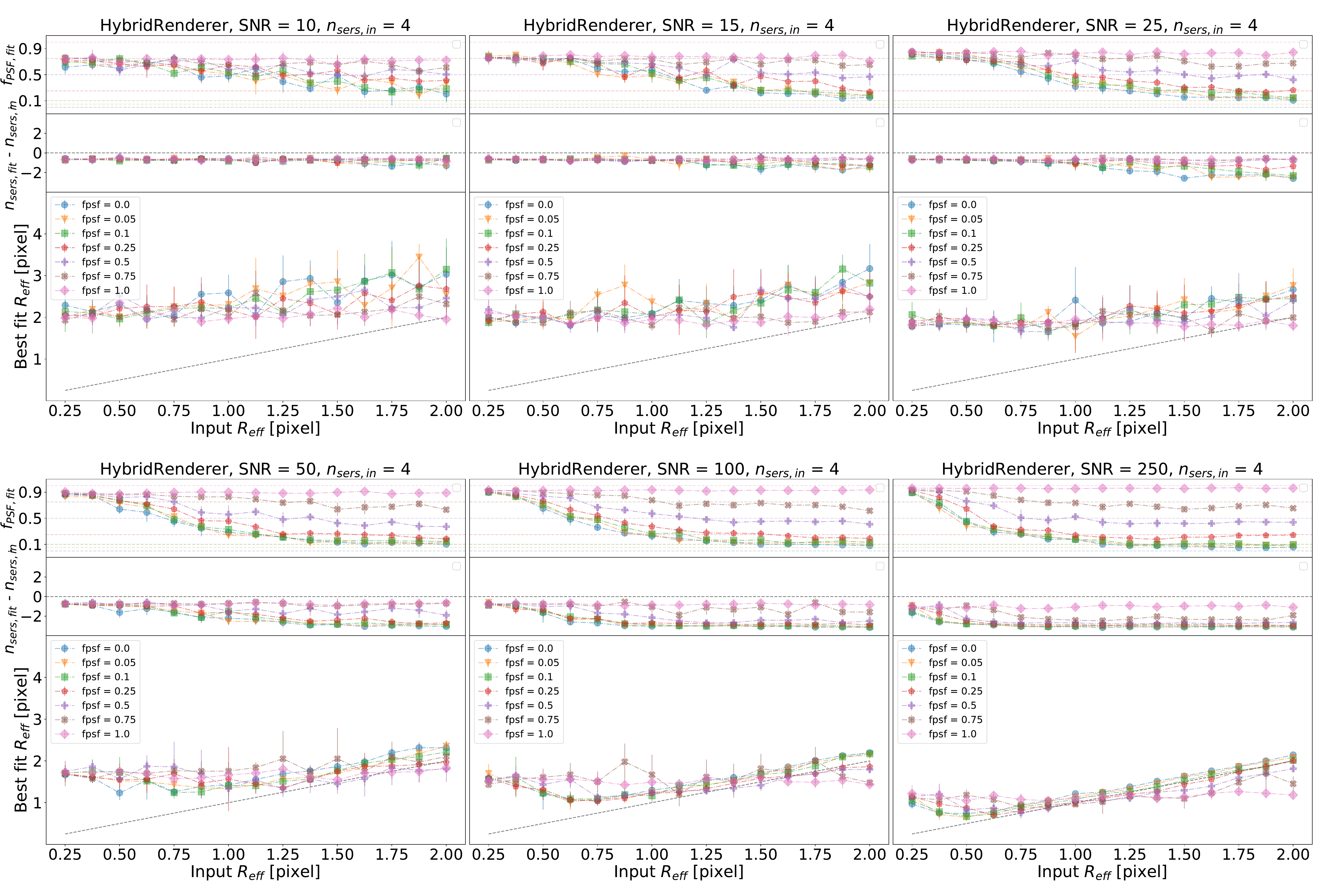}
    \caption{Two-component decomposition can determine the fraction of a target's flux that is contained in unresolved PSF-like component for a wide range of SNR, however it cannot estimate morphological properties for underlying the resolved stellar distribution.
    Here we present results from pysersic two-component decompositions for our suite of mock LRDs at varying SNR with $n_{\textrm{sers,in}}=4$. We find that pysersic has difficulty simultaneously recovering all of the input parameter values over the full SNR range, but the difficulty gets exacerbated at low-SNR. For mock objects with $\textrm{SNR} \lesssim 50$, best-fit $R_{\textrm{eff}}$ measurements for the \sersic component do not reflect the size of the input \sersic profile that was used to construct each given simulated object. We also find that the posterior distributions for the best-fit $n_{\textrm{sers}}$ are unconstrained for models of all sizes across the full SNR range. Lastly, we find that for galaxies with small $R_{\textrm{eff,in}}$, pysersic has a tendancy to overestimate $f_{\textrm{PSF}}$. This deviation between $f_{\textrm{PSF, in}}$ and $f_{\textrm{PSF,fit}}$ happens at increasingly larger $R_{\textrm{eff,in}}$ for lower SNR targets.}
   \label{fig:pysersic_hyb_ps+sers_n4}
\end{figure*}

Morphologies of real galaxies are often too complex to be modeled using a single \sersic component. It is common to decompose the morphologies of more typical AGN host galaxies at low-$z$ into two-component models that separate the central PSF-like emission from an underlying stellar population that can be modeled with a \sersic profile. An aim of ours in this work is to determine under which sets of assumptions this two-component decomposition might be feasible for existing observations of LRDs. We present results for our pysersic HybridRenderer two-component decomposition for the subset of our mock LRDs whose stellar populations are modeled as $n_{\textrm{sers,in}} = 4$ in Figure \ref{fig:pysersic_hyb_ps+sers_n4}. The fits to mock LRDs created from differing $n_{\textrm{sers,in}}$ values are given in the appendix. Figure \ref{fig:pysersic_hyb_ps+sers_n4} is comprised of 6 main panels, each representing a different object SNR. Within each of these panels, we show the best-fit $f_{\textrm{PSF}}$ (A), $n_{\textrm{sers,fit}}$-$n_{\textrm{sers,in}}$ (B), and $R_{\textrm{eff,fit}}$ (C) as a function of $R_{\textrm{eff,in}}$.  For all mock LRDs with $\textrm{SNR} \lesssim 50$, we find that pysersic $R_{\textrm{eff,fit}}$ measurements fail to recover input $R_{\textrm{eff,in}}$ values across the full range of $R_{\textrm{eff,in}}$ we explored. Pysersic has slightly more success with $\textrm{SNR}=100,250$ objects, but only for those with $R_{\textrm{eff,in}} \gtrsim 1.325$ and $R_{\textrm{eff,in}} \gtrsim 0.75$ pixels, respectively. Looking towards subpanel (B), it is also clear that pysersic has difficulty recovering $n_{\textrm{sers,in}}$. The posterior distributions for individual fits to mock images spanning the full range of input parameter values reveal that $n_{\textrm{sers}}$ is completely unconstrained.

Although pysersic  two-component decomposition is largely unreliable for our mock-LRDs, subpanels (A) of Figure \ref{fig:pysersic_hyb_ps+sers_n4} reveal that it can still be somewhat useful in determining relatively how PSF-like LRDs might be. Across the full SNR range explored, objects constructed from models with large $R_{\textrm{eff,in}}$ and low $f_{\textrm{PSF,in}}$ values are consistently fit with smaller $f_{\textrm{PSF,fit}}$ values than their more intrinsically compact counterparts, although this is seen more strongly in objects with higher-SNR. In other words, pysersic deems that an object with $f_{\textrm{PSF,in}}\sim 1$ is nearly indistinguishable from another with $f_{\textrm{PSF,in}}\sim 0$ but with $R_{\textrm{eff,in}} \sim 0.5$; this makes sense considering that they are both intrinsically compact objects that would be considered to be PSF-dominated when convolved with the instrumental PSF. The usefulness with the two-component decomposition here rather than a basic \sersic fit is that the best fit $f_{\textrm{PSF,fit}}$ values are less degenerate across a wider range of SNR than $R_{\textrm{eff,fit}}$ is with the one component fit. We believe that this is especially important for low-SNR targets where the wings of an underlying extended \sersic profile might easily be lost in the noise; it would be easier for pysersic to determine that a certain fraction of light is within the PSF than it would be for it to make an accurate measurement on the underlying $R_{\textrm{eff}}$. It allows for a spectrum to exist between a source being classified as ``extended'' or ``PSF-dominated'' as opposed to a binary; all without over-interpreting the underlying physical properties of the source. The results for GALFIT are conistent with this picture, although with some more scatter.

\subsubsection{PSF fits}
\begin{figure*}
    \centering
    \includegraphics[width=1\linewidth]{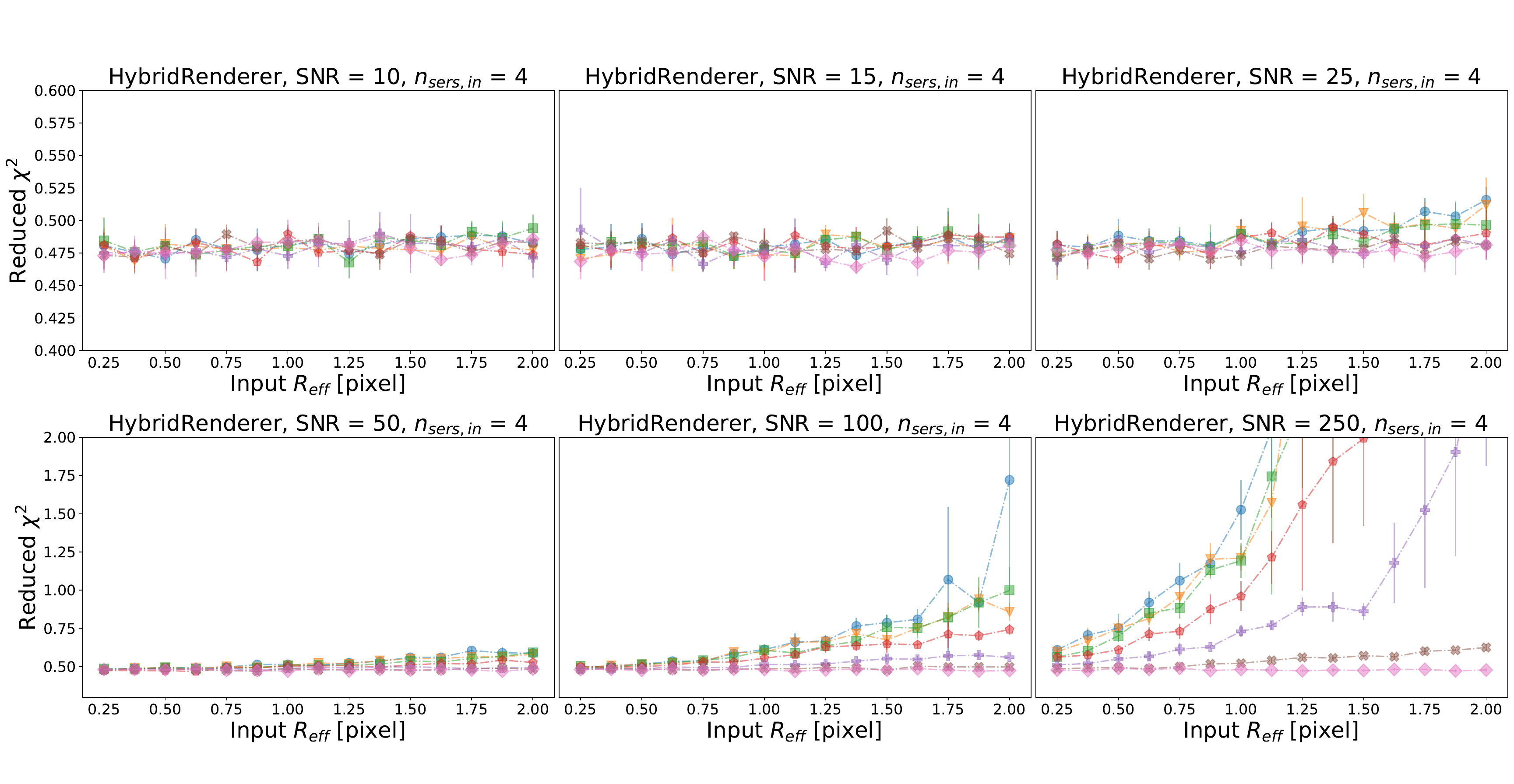}
    \caption{All simulated LRD models with $\text{SNR} \lesssim 25$ are indistinguishable from point sources, regardless of their intrinsic compactness. Here we present reduced-$\chi^{2}$ values for pysersic PSF HybridRender fits for suite of mock LRD cutouts. For objects with $\textrm{SNR} \lesssim 25$, all images are fit reasonably well by a single PSF component for all $f_{\textrm{PSF,in}}$ and $R_{\textrm{eff,in}}$ values, meaning that it is difficult to ascertain whether or not an object is intrinsically extended within the limits of the background noise. When $\textrm{SNR} \gtrsim 50$, profiles with low-$f_{\textrm{PSF,in}}$ and large-$R_{\textrm{eff,in}}$ are no longer fit well by a single PSF; they can be definitively classified as extended.}
   \label{fig:pysersic_PSFfit}
\end{figure*}

The final question we explored with regards to the limitations of morphological fitting was: under what set of model assumptions would a subset of mock LRDs be distinguishable from a pure PSF component. We addressed this by fitting our synthetic images with a model PSF that matches the one used to construct them, and then by measuring the reduced-$\chi^{2}$ values for the fit. We similarly did this test with both pysersic and GALFIT.

The results for the pysersic HybridRenderer fitting are presented in Figure \ref{fig:pysersic_PSFfit}.  We note that the baseline reduced-$\chi^{2}$ values for these fits are $\sim 0.5$ instead of the expected unity. This is due to an underestimation of noise that arises from images with correlated noise being fit within the per-pixel uncertainties characterized by an uncorrelated RMS image. The results for fits performed by pysersic/PixelRenderer and GALFIT are included in the Appendix

All cutouts with $\textrm{SNR} \lesssim 25$ are fit reasonably well by a basic PSF model. The average reduced-$\chi^{2}$ values for the fits are indistinguishable from one another. Starting at around $\textrm{SNR} \gtrsim 50$, more intrinsically extendeded cutouts with large-$R_{\textrm{eff,in}}$ and/or small-$f_{\textrm{PSF,in}}$ have higher average reduced-$\chi^{2}$ than their more compact counterparts. This discrepancy continues to increase as a function of SNR, and is seen in both the fits performed with pysersic/PixelRenderer and GALFIT. This result implies that within a certain noise threshold, it is difficult to determine whether or not an object is extended. Within the context of LRDs, this highlights that just because an object can be fit by a model PSF with little resulting residuals does not mean that it is inherently PSF-like within the $\textrm{SNR} \lesssim 25$ range.

\section{Simulations in the context of observed LRD populations}
Throughout this work, we have explored the capabilities of the popular morphological fitting codes pysersic and GALFIT with respect to recovering the input parameter values used to construct a simulated suite of potential LRD-like model galaxies. In this section, we contextualize these results with respect to the observed population of LRDs presented in \citet{koko24census}.

As mentioned in Section 2, we worked with cutouts from deep F444W images of \textit{JWST} blank fields that were publicly available and fully reduced on the DJA. These include data from the following programs: CEERS(\# 1345; PI: S. Finkelstein; \citealt{bagl23ceersNIRCAM,fink23ceers}), PRIMER (\# 1837; PI: J. Dunlop), FRESCO (\# 1895; PI: P. Oesch; \citealt{oesc23fresco}), JADES (\# 1180, 1210, 1286, and 1287; PIs: D. Eisenstein and N. Luetzgendorf; \citealt{eise23jades,eise23data-relase-jades}), and JEMS (\# 1963; PI: C. Williams).

\subsection{The SNRs of observed LRDs}

As shown in the previous sections, there is a SNR limit at which morphological fitting codes become less reliable. Contextualizing the results of the simulations requires us to fully understand the morphological measurements that have been done on LRD cutouts from real observations. As mentioned in Section 2, we similarly analyze  3" cutouts from the DJA mosaics described in \citet{koko24census}. We also sort this LRD samble into SNR bins that match those used in our analysis of our mock LRDs. We compute the SNR for each LRD following the method we describe in Section \ref{subsubsec:noise} to account for correlated noise and ensure that we are consistent with how we define SNR for our mock sample.

\subsection{Fitting the morphologies of observed LRDs}

\begin{sidewaystable*}[p]
\centering
\footnotesize
\vspace{18cm}
\begin{tabular}{ccccccccccccccc}
ID & R.A. & Dec. & $R_{\textrm{eff}}$  & $\sigma_{R_{\textrm{eff}}}$ & $n_{\textrm{sers}}$  & $\sigma_{n_{sers}}$  & $\chi ^{2}_{\nu}$ & $R_{\textrm{eff}}$ & $\sigma_{R_{\textrm{eff}}}$ & $n_{\textrm{sers}}$ & $\sigma_{n_{sers}}$ & $f_{\textrm{PSF}}$ & $\sigma_{f_{\textrm{PSF}}}$ & $\chi ^{2}_{\nu}$  \\
 & deg & deg &[px] (\sersic) & [px] (\sersic) & (\sersic) & (\sersic) & (\sersic) & [px] (\sersic+ps) & [px] (\sersic+ps) & (\sersic+ps) & (\sersic+ps) &  & & (\sersic+ps) \\
 (1) & (2) & (3) & (4) & (5) & (6) & (7) & (8) & (9) & (10) & (11) & (12) &  (13) & (14) & (15) \\
\hline
1498 & 34.3722 & -5.21 & 0.344 & 0.038 & 1.081 & 0.14 & 0.4344 & 0.812 & 0.325 & 1.289 & 0.338 & 0.844 & 0.129 & 0.4375 \\
2981 & 34.441 & -5.2093 & 0.495 & 0.062 & 1.099 & 0.239 & 0.5939 & 1.145 & 0.677 & 1.773 & 0.876 & 0.799 & 0.167 & 0.593 \\
3099 & 34.2605 & -5.2092 & 0.367 & 0.053 & 1.163 & 0.233 & 0.4023 & 1.235 & 0.966 & 1.941 & 0.959 & 0.854 & 0.117 & 0.4045 \\
5434 & 34.3774 & -5.209 & 0.637 & 0.028 & 3.388 & 0.27 & 0.5843 & 1.084 & 0.134 & 3.918 & 0.498 & 0.359 & 0.053 & 0.58 \\
14381 & 34.4031 & -5.2082 & 0.87 & 0.497 & 4.059 & 1.492 & 0.3969 & 2.079 & 1.391 & 3.585 & 1.538 & 0.712 & 0.21 & 0.3968 \\
14605 & 34.3673 & -5.2042 & 0.628 & 0.073 & 5.3 & 0.341 & 0.465 & 4.322 & 0.324 & 4.347 & 0.92 & 0.559 & 0.036 & 0.4446 \\
16053 & 34.2937 & -5.1965 & 0.395 & 0.131 & 4.002 & 1.554 & 0.4277 & 1.921 & 1.361 & 3.511 & 1.53 & 0.835 & 0.144 & 0.4269 \\
18537 & 34.4654 & -5.1953 & 0.612 & 0.053 & 1.228 & 0.255 & 0.4768 & 0.828 & 0.159 & 2.168 & 0.982 & 0.476 & 0.092 & 0.4793 \\
22388 & 34.4629 & -5.192 & 0.291 & 0.012 & 3.078 & 0.068 & 8.8309 & 2.413 & 2.211 & 3.059 & 2.389 & 0.768 & 0.05 & 8.4343 \\
23868 & 34.3992 & -5.1843 & 0.377 & 0.094 & 2.445 & 1.186 & 0.3887 & 1.669 & 1.217 & 3.198 & 1.533 & 0.809 & 0.17 & 0.3891 \\
29303 & 34.4319 & -5.1828 & 0.474 & 0.058 & 3.695 & 0.681 & 0.4112 & 1.402 & 0.402 & 2.456 & 1.12 & 0.59 & 0.093 & 0.4094 \\
32147 & 34.5081 & -5.1801 & 0.443 & 0.027 & 1.1 & 0.126 & 0.5435 & 3.873 & 0.48 & 3.837 & 1.016 & 0.859 & 0.017 & 0.5345 \\
32402 & 34.3635 & -5.177 & 0.418 & 0.064 & 1.043 & 0.225 & 0.541 & 1.159 & 0.84 & 1.549 & 0.668 & 0.794 & 0.218 & 0.5422 \\
35367 & 34.2689 & -5.1767 & 0.32 & 0.022 & 0.917 & 0.05 & 0.9073 & 1.086 & 0.998 & 1.481 & 0.507 & 0.883 & 0.112 & 0.9531 \\
37563 & 34.4608 & -5.1758 & 0.592 & 0.042 & 5.592 & 0.317 & 0.4782 & 4.305 & 0.193 & 5.535 & 0.344 & 0.492 & 0.023 & 0.4678 \\
39509 & 34.363 & -5.1732 & 0.371 & 0.093 & 1.372 & 0.463 & 0.4392 & 1.315 & 0.971 & 2.195 & 1.218 & 0.806 & 0.178 & 0.4399 \\
41129 & 34.439 & -5.1705 & 0.394 & 0.035 & 1.008 & 0.107 & 0.5276 & 1.007 & 0.684 & 1.384 & 0.354 & 0.838 & 0.159 & 0.5298 \\
44531 & 34.408 & -5.1691 & 0.415 & 0.082 & 4.489 & 1.143 & 0.374 & 1.893 & 1.118 & 3.545 & 1.508 & 0.753 & 0.137 & 0.3736 \\
45202 & 34.3462 & -5.1682 & 0.482 & 0.033 & 1.083 & 0.118 & 0.4854 & 0.837 & 0.144 & 1.039 & 0.163 & 0.621 & 0.097 & 0.4879 \\
46221 & 34.4429 & -5.1682 & 0.416 & 0.026 & 4.848 & 0.342 & 0.5157 & 3.449 & 0.438 & 1.711 & 0.517 & 0.737 & 0.027 & 0.5172 \\
46281 & 34.3316 & -5.1592 & 0.398 & 0.06 & 1.175 & 0.294 & 0.3785 & 1.193 & 0.831 & 1.984 & 1.123 & 0.819 & 0.159 & 0.3796 \\
\end{tabular}
\caption{A summary of the pysersic morphological fitting performed on the F444W LRD cutouts. Columns (4) through (8) represent the results from the single \sersic fitting, and columns (9) through (15) for the two-component decomposition. A full version of this table is available on \href{https://github.com/ke27whal/LRD-Morphology/blob/main/whalen_et_al_fit_table.fits}{github}. A value of -99 means that the fit failed for that particular target.\vspace{9cm}}
\label{tab:fit_results}
\end{sidewaystable*}

A major reason why we adopted the sample of LRDs in \citet{koko24census} is that it is one of the few works in the literature that has presented morphological measurements for a statistically significant sample. There have been some attempts at parametric morphology fitting elsewhere, including for several individual LRDs/small samples \citep[e.g.,][]{naid25black_hole_star,wang24broadline_outflow_lrd,rona25megaLRD,tayl25capers}. Some other works have made attempts at selecting unresolved targets by comparing their sizes to a stellar-locus that has been fit with forward-moldeled galaxy profiles \citep[e.g.,][]{akin24overabundance_lrd,hvid25_rubies-bl-lrds}. \citet{koko24census} present their best fit half-light radii as size upper limits, although the other best fits to ancillary morphological parameters such as $n_{\textrm{sers}}$ are not included. Since the primary analysis in this work has also explored the covariance between underlying $n_{\textrm{sers}}$ and $R_{\textrm{eff}}$, we have decided to refit the F444W morphologies of the \citet{koko24census} cutouts using pysersic to ensure that we have best fits to all of model parameters. We have additionally performed a two-component decomposition. We constructed RMS images for all of our fits that included source Poisson noise following the documentation on the DJA imaging products web page \footnote{https://dawn-cph.github.io/dja/blog/2023/07/18/image-data-products/}.

We include a summary of the results to our fitting in Table \ref{tab:fit_results}. Our priors are the same as described in Section 3. Similar to \citet{koko24census}, we primarily measured that the LRDs have mean $R_{\textrm{eff}} \lesssim 1$ pixel. However, our sizes were slightly larger and had more scatter. To ensure that our fits were comparable to those from \citet{koko24census}, we performed a second \sersic fit to the data but instead of using the uniform prior on $R_{\textrm{eff}}$ that was detailed in \citet{koko24census}, we used a Gaussian prior where the mean was the \citet{koko24census} best fit value and the width was 0.5 pixels. We were able to recover the best-fit \citet{koko24census} measurements using this Gaussian prior. We also found that the $\chi ^{2} _{\nu}$ values were comparable for both priors, suggesting that both fits describe the data equally well. The parameter space is likely highly degenerate and it is possible that there are many local minima in the likelihood function. We continue to quote the results from the uniform prior fits since those are less biased towards a particular value and match the assumptions for the only other reported morphological fitting of this sample.

\subsection{Exploring the compactness of observed LRDs}
Photometrically selected LRDs are by design selected to be morphologically compact. They must satisfy a ``compactness criteria'' where the flux ratio between $0\farcs04$ and $0\farcs02$ radius apertures is $< 1.7$. Here, we compare the measured compactness ratio to other morphological properties measured using pysersic to determine how well it traces underlying morpholgical structure. 
\subsubsection{\sersic fit}

\begin{figure*}
   \includegraphics[width=1\linewidth]{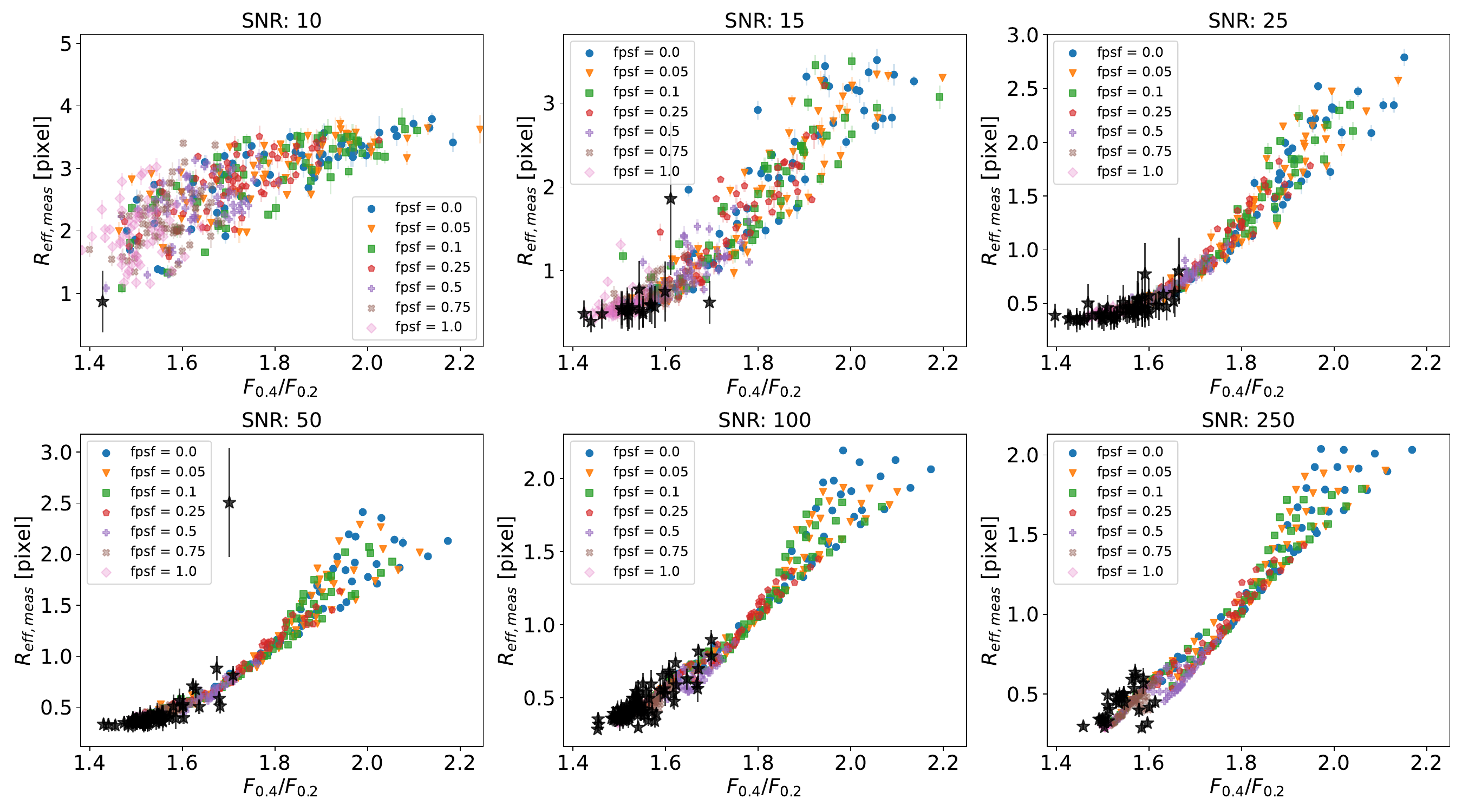}
    \caption{Observed LRDs occupy the same region of $R_{\textrm{eff,fit}}$--$F_{0\farcs04}/F_{0\farcs02}$ parameter space as simulated LRDs with diverse sets of input morphological parameters. Each panel depicts the measured $R_{\textrm{eff}}$ value versus compactness ratio ($F_{0\farcs04}/F_{0\farcs02}$) for observed LRDs (black stars) and our suite of simulated LRD-like objects (multi-colored, color corresponding to input $f_{\textrm{PSF}}$ for a given SNR bin. For the simulated LRDs, $R_{\textrm{eff,in}}$ scales with compactness ratio with this relationship tightening with SNR. Each color point represents different $f_{\textrm{PSF,in}}$, but spans all models regardless of \sersic $R_{\textrm{eff,in}}$ or $n_{\textrm{sers,in}}$, as evident by the different ``tracks'' that emerge as SNR increases. Observed LRDs occupy the most compact (Lower-left) region of this parameter space.}
   \label{fig:reff_vs_compact} 
\end{figure*}
We begin by comparing the results of our single component \sersic fits to those measured for our suite of mock LRDs. This is shown in Figure \ref{fig:reff_vs_compact}. Each of the panels represents a different SNR bin. The fits to observed LRDs are depicted with black stars and the results to our mock sample are given as multicolor points whose color corresponds to a particular input $f_{\textrm{PSF}}$. We find that the measured $R_{\textrm{eff}}$ for simulated LRDs scales with compactness ratio ($F_{0\farcs04}/F_{0\farcs02}$). This correlation also tightens with increasing SNR. We note that the simulated points here encompass models that span the full range of input $R_{\textrm{eff}}$ and $n_{\textrm{sers}}$; this becomes clear in high SNR panels with the emergence of distinct tracks. By design, the observed LRDs occupy the bottom left, or most compact regions of this parameter space.


For each of the observed LRDs in this sample, we selected its 5 nearest neighbors in $R_{\textrm{eff,fit}}$---$F_{0\farcs04}/F_{0\farcs02}$ space within its SNR bin and computed the weighted average and variance of $R_{\textrm{eff,in}}$, $n_{\textrm{sers,in}}$, and $f_{\textrm{PSF,in}}$ values to estimate which models best described the observed LRDs. Using the means and standard deviations of each LRD's nearest neighbors, we performed a Monte Carlo simulation where we drew 5,000 samples from the corresponding Gaussian distributions to estimate the covariance between model input parameters for each SNR bin. We unsurprisingly find covariance between $R_{\textrm{eff,in}}$ and $f_{\textrm{PSF,in}}$ across all SNR bins. In other words, models with more extended \sersic components require higher $f_{\textrm{PSF}}$ to satisfy the LRD compactness criteria. This covariance grows stronger with increasing SNR. We also find that the average nearest neighbor input parameter measurements have large uncertainties. This likely is partly due to the pysersic fits to modeled LRDs having large uncertainties at low to moderate SNR, and partly due to an underlying morphological diversity within the LRD sample.

\subsubsection{\sersic + PSF fits}
\begin{figure*}
    \centering
    \includegraphics[width=1\linewidth]{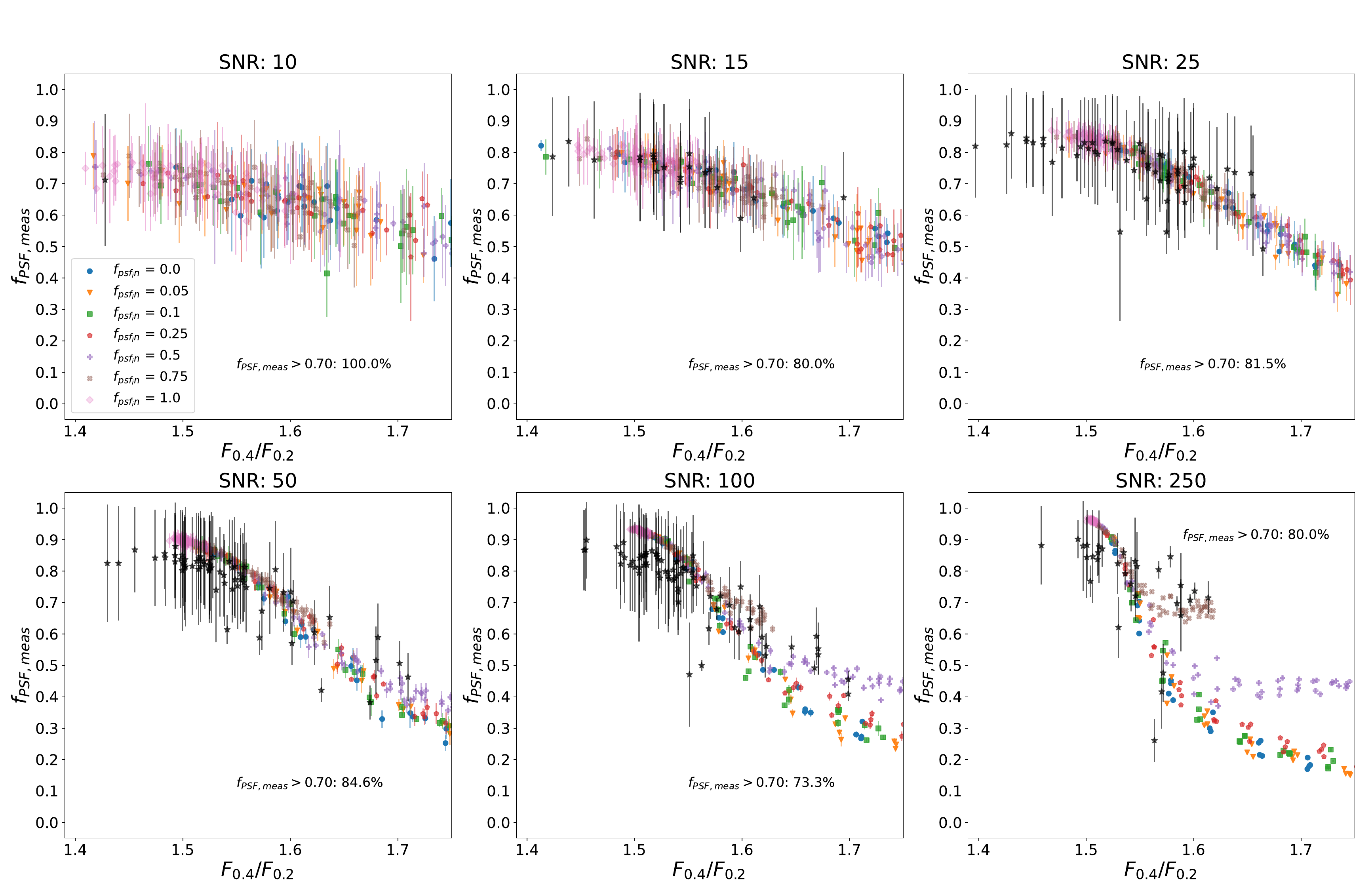}
    \caption{Majority of observed LRDs are dominated by unresolved, PSF-like emission across a wide range of SNR. However, $\sim15\%$ of LRDs that are compact enough to be selected exhibit evidence of having significant resolved F444W emission. Here we present results from our two-component decomposition using pysersic on observed and mock LRDs. Each panel represents a different $\text{SNR}$ bin and shows the relationship between $f_{\textrm{PSF,fit}}$ and $F_{0\farcs04}/{F_{0\farcs02}}$, or the compactness ratio. Best fits to observed LRDs are depicted with black stars and those for modeled LRDs are given as multi-colored points with each color representing the input $f_{\textrm{PSF,in}}$ value used to construct each given model.}   \label{fig:fpsf_vs_compact}
\end{figure*}

In \ref{ssec:decomp}, we establish that two-component \sersic/PSF decomposition could be an invaluable tool for estimating the fraction of unresolved emission in LRDs. We test this by performing a two-competent fit to the observed LRD cutouts and comparing the results for those of our simulated LRDs. This is shown in Figure \ref{fig:fpsf_vs_compact}; the black stars represent fits to observed LRDs, and the multi-colored points are fits to our mock sample of LRD-like objects. We confirm that $f_{\textrm{PSF,fit}}$ correlates with the compactness ratio for both the real and simulated LRDs across every SNR bin. However, the scatter around this relationship is large for $\text{SNR} \lesssim 25$, making it difficult to differentiate between objects that are mostly unresolved versus those that are compact yet have a detectable extended component. 

This trend becomes much more pronounced for objects with $\text{SNR} \gtrsim 50$. The $\text{SNR} \sim 50$ bin shows that simulated LRDs are more likely to have $f_{\textrm{PSF,fit}} \lesssim 0.7$ when  $F_{0".4}/F_{0".2} \gtrsim 1.6$. This smooth trend between $f_{\textrm{PSF,fit}}$ and compactness ratio begins to resemble an increasingly steep step-function for objects with $\text{SNR} \gtrsim 100$. For $\text{SNR} \sim 100$, simulated LRDs with $F_{0".4}/F_{0".2} \gtrsim 1.65$ almost entirely are best fit with $f_{\textrm{PSF,fit}} < 0.5$. This step function grows even more stark at $\text{SNR} \sim 250$, where  simulated LRDs with $F_{0".4}/F_{0".2} \gtrsim 1.6$ have  $f_{\textrm{PSF,fit}} < 0.3.$. 

Our best-fit $f_{\textrm{PSF,fit}}$ values for the observed LRDs roughly trace the fits to our models in every SNR bin. The vast majority of the observed LRDs are consistent with having most of their F444W emission arise from an unresolved, PSF component, as evident by $\sim 75 \%$ of the observed LRD sample being best-fit by models with $f_{\textrm{PSF,fit}} > 0.7 $. This fraction of PSF-dominated observed LRDs is given for each SNR bin in Figure \ref{fig:fpsf_vs_compact}, as well. This evidence highlights that most observed LRDs are compact enough to be morphologically consistent with being completely unresolved, meaning that their F444W morphology does not have much contribution from a potential host galaxy. However, the other 25\% reveal a potentially interesting subpopulation that is still highly compact and is mostly unresolved. However, it is possible to potentially detect extended emission from their host galaxies, especially for LRDs with $\text{SNR} \gtrsim 50$. This gradient of morphological diversity highlights that it is possible that not all LRDs arise from the same physical phenomena, or at the very least that they might not be in the same phase of their evolution.

\subsubsection{Non-parametric Compactness Indicators}
\begin{figure*}
    \includegraphics[width=0.9\linewidth]{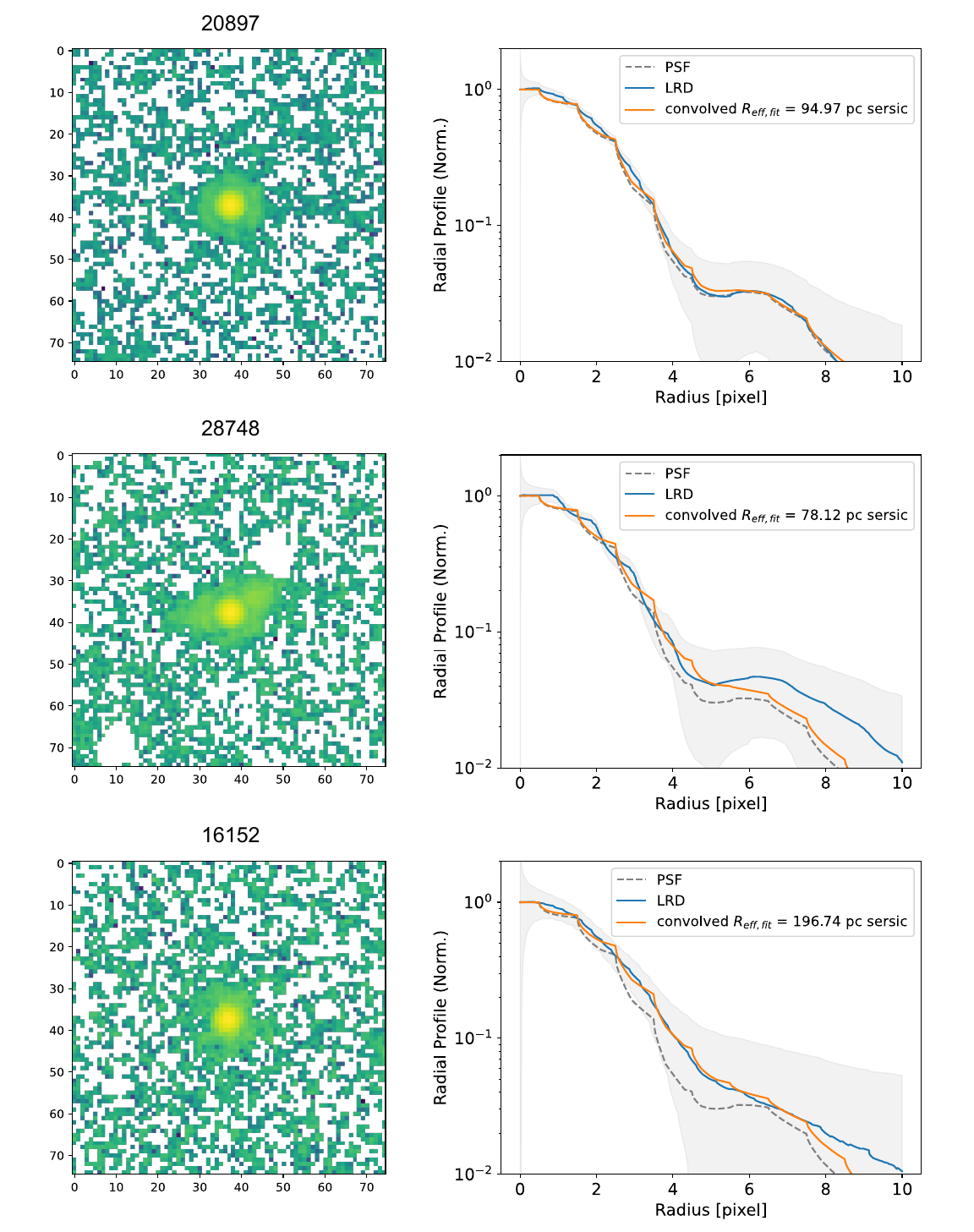}
    \caption{There is morphological diversity in the full population of observed LRDs. We present F444W cutouts of three selected LRDs \textit{(left)} along with their corresponding radial profiles \textit{(right)}. All of the cutouts have a logarithmic stretch, and we show radial profiles for each LRD's best-fit single component convolved \sersic profile (orange) and the empirical PSF (grey, dashed). We select these three to highlight the range in LRD morphologies that exist. LRD 20897 is PSF-dominated; it was best fit  by profiles with $R_{\textrm{eff,fit}} \sim 95$ pc (one component \sersic fit), or $f_{\textrm{PSF,fit}}\sim 0.85$ (two component decomposition). LRD 28748 is centrally compact, but exhibits diffuse wings that aren't recovered in the best-fit \sersic; it was best fit with $R_{\textrm{eff,fit}} \sim 78$ pc or $f_{\textrm{PSF,fit}}\sim 0.70$. LRD 16152 is more extended than the PSF at all scales; it was best fit with $R_{\textrm{eff,fit}} \sim 196$ pc or $f_{\textrm{PSF,fit}}\sim 0.53$.}
    \label{fig:cutouts_profiles}
\end{figure*}
Two-component decomposition fitting for LRDs reveals that although all LRDs are morphologically compact, some have detectable amounts of resolved emission in the F444W band. Here we visually inspect the 1-dimensional radial profiles as measured from the LRD cutouts to compare their extent versus that of the empirical PSF of the field in which each LRD was detected and the PSF-convolved best-fit \sersic profile as measured by pysersic. In  Figure \ref{fig:cutouts_profiles} we present F444W cutouts (\textit{left}) and radial profiles (\textit{right}) for three LRDs (IDs: 20897, 28748, 16152) that are representative of the morphological diversity of the LRD population. For the radial profile plots, the profile of the observed LRD is shown in blue, the convolved best-fit \sersic in orange, and the PSF in grey. 

The LRDs are shown in order of increasing extent. The radial profile for LRD 20897 closely matches that of the PSF, suggesting that the vast maority of F444W emission for this source is unresolved, or PSF-line. This is consistent with the best parametric fits to this source; single component \sersic fitting suggests $R_{\textrm{eff,fit}} \sim 0.35 \text{ px} \sim 95 \text{ pc}$ and two-component decomposition gives $f_{\textrm{PSF,fit}} \sim 0.85$. This source is representative of majority of the LRD population. Next, in the middle row of \ref{fig:cutouts_profiles}, we present LRD 28748. This source is also centrally compact; its single component \sersic fit measures $R_{\textrm{eff,fit}} \sim 0.296 \text{ px} \sim 78 \text{ pc}$ and two-component suggests $f_{\textrm{PSF,fit}} \sim 0.70$. This is also evident in how the source's radial profile closely traces the PSF and the best-fit \sersic out to $R\sim5$ px, but then deviates to highlight that more diffuse, larger scale extended emission. The nature of this emission is currently unknown, it could indicate that this source is perhaps an interacting pair of galaxies, or it could simply be a low-$z$ interloper that was centered on the compact emission by chance. It is worth mentioning that \citet{chen25LRDoffcenter} performed SED fitting for three different LRDs with off-center, extended emission and determined that the extended blobs were associated with their central cores. The last source shown here is LRD 16152. This is a target that is more extended than the PSF at all scales, and closely matches the best fit $R_{\textrm{eff,fit}} \sim 0.82 \text{ px} \sim 197 \text{ pc}$ \sersic. Two component decomposition also suggests that only 50\% of the light from this source is contained in a PSF component. This source is similar to $\sim15\%$ of the total LRD population, and is more likely to have a detectable host component than the LRDs whose profiles more closely resemple the PSF.

This comparison we have done here is similar to the work presented in \citet{diam12CSstarform} and \citet{sell14CSoutflow} for a population of $z\sim0.5$ massive ($10^{10} \ M_{\odot}$), extremely compact starburst galaxies. The sample of galaxies presented there is interesting in the context of LRDs. They were originally selected from the Sloan Digital Sky Survey (SDSS) quasar catalog for being bright, blue, and unresolved. However, they were reclassified as starbursting or as very young post-starburst galaxies due to their relatively weak nebular emission and extremely high mid-IR derived SFRs ($> 300 \text{ M}_{\odot}/\text{yr}$; \citealt{trem07CSoutflow}). Their SEDs are characterized by a blue rest-frame UV continuum that leads into a rising red/near-IR slope, qualitatively similar to that of LRDs. They are not detected in the X-ray, nor do they have radio signatures of hosting AGN \citep[e.g.][]{sell14CSoutflow,pett20radio}. 

The most striking similarity between this population of $z\sim0.5$ compact starburst and LRDs is their extreme compactness. The $z\sim$ 0.5 compact starbursts are completely unresolved in SDSS. Follow-up Hubble Space Telescope observations reveal that most of them are dominated by an unresolved core (or two cores in close proximity) embedded in faint, tidal features indicative of having recently undergone a major merger \citep[e.g.,][]{diam12CSstarform,diam21CSoutflows,sell14CSoutflow}. Best-fits as determined with GALFIT reveal that these sources have extremely compact ($R_{Rff}\sim 100$ pc) morphologies, similar to the sizes we recover for a portion of LRDs. We also visually inspected the F1115W cutouts for the full sample of LRDs. The F115W band has a significantly smaller PSF and their DJA mosaics are more finely sampled than the long-wavelength bands, allowing us to resolve more compact features. Several of the LRDs with more extended F444W morphologies appear to have either dual-cores or features that resemble tidal tails in the F1115W images, similar to what we see in the $z\sim0.5$ compact starbursts. The presence of disturbed LRD short-wavelength morphologies has been noted by \citet{rina24UV-LRDs}; many of these extended rest-UV LRDs also exhibited broad H$\alpha$ with FWHM $1200-2900$ km/s. Emission line diagnostics put these sources in the ``composite'' range of the diagnostic ``OHNO'' diagram \citep[e.g.,][]{trou11high-z-bpt,trum23_highz-emline-diag}. \citet{rina24UV-LRDs} has similarly interpreted these UV morphologies as indication that some LRDs might be the result of interacting galaxies triggering star formation and/or AGN activity.

A major difference between LRDs and $z\sim0.5$ compact starbursts is that up to $\sim80\%$ of LRDs exhibit broadened Balmer lines \citep[e.g.,][]{green24uncover_broadline_lrd,hvid25_rubies-bl-lrds}, while the $z\sim0.5$ compact starbursts lack them although they do show evidence for high-velocity ($> 1000$ km/s; \citealt{trem07CSoutflow,davi23hizea_scaling}), large-scale ($\sim$ tens of kpc; e.g., \citealt{geac13CSmolec,geac14CSstellarfeedback,geac18molecoutflow,rupk19makani,rupk23makani_wind}) Mg II, [OII], and molecular gas outflows. \citet{bagg24starburst_broadline_lrd} suggested that the presence of broad lines in LRDs does not necessarily require an AGN, and that galaxies with extreme stellar densities could be kinematically consistent with observations. This sample of  $z\sim0.5$ compact starburst galaxies are referenced in \citet{bagg24starburst_broadline_lrd} and are shown to be dense enough to match the physical requirements facilitated by their calculation. We stress that a vast majority of LRDS are most consistent with the AGN interpretation based on their spectra and their morphologies. However, there is a significant minority that could be analogous with these compact starbursts seen at $z\sim 0.5$, or composites between a strong, central starburst and an AGN. This sub-sample is particularly interesting in understanding how the AGN-starburst connection unfolds at high-$z$.

\section{Summary \& Conclusions}
The discovery of LRDs in \textit{JWST} blank fields have challenged our understanding of galaxy and supermassive black hole formation. Their multiwavelength properties are confounding; they are morphologically compact and have broad emission lines as would be expected for an AGN, but their unusual SED shape and lack of detection in stacked archival X-ray images does not quite fit our canonical understanding of AGN. Being able to characterize the physical nature of LRDs first requires us to find the limitations of the tools we use to measure their properties.

Parametric morphological fitting has long been used to estimate the sizes and shapes galaxy surface brightness profiles. This technique fits multi-parameter models that have been convolved with the instrumental PSF to image cutouts, describing their underlying morphology. One challenge in measuring the sizes of LRDs is their extreme compactness; they appear to be point like in images, suggesting that they are almost entirely unresolved. First attempts at measuring their morphologies using parametric fitting codes like pysersic or GALFIT reveal that their best-fit F444W radii span $30 \lesssim R_{\textrm{eff}}/\text{pc} \lesssim 200$. This corresponds to image-scale sizes of $\lesssim 1 \text{ pixel}$ (before considering convolution with the PSF). This type of extreme compactness pushes against the computational limits of parametric morphological fitting, as the fitting becomes even more sensitive to the user-selected model PSF and as well as the level of background noise.

A main objective of this work was to address the limitations of parametric morphological fitting tools in estimating the compact sizes of LRDs. In particular, we were interested in determining if there was a minimum intrinsic size that would be recoverable, as well as if there was a SNR dependence for the accuracy of these fitting codes. We examined this question by constructing a suite of simulated objects to be fit. We made the simple assumption that LRDs could be modeled as the linear combination of an ``extended'' \sersic profile (parameterized by $R_{\textrm{eff,in}}$ and $n_{\textrm{sers}}$) and an embedded PSF component, where their relative contribution is scaled by parameter $f_{\textrm{PSF}}$, or the fraction of F444W flux contained in the PSF component. We simulated image cutouts for objects with different parameter permutations in the range $0.25 < R_{\textrm{eff}}/\text{pixel} < 2$, $1 < n < 4$, $0 < f_{\textrm{PSF}}< 1$, and  varied the SNR in six bins within $10 < \text{SNR} < 250$. We then performed fitting with pysersic and GALFIT to examine if there exists a region of parameter space at which the results become unreliable. We fit three models to our simulated images: 1) a single component \sersic, 2) two-component \sersic + PSF, and 3) just the PSF. The results of this analysis can be summarized as such:
\begin{itemize}
    \item Both, pysersic and GALFIT best recover the input parameter values used to construct the simulated images at high-SNR for the single-\sersic fit. The fits for cutouts with $\text{SNR} \lesssim 15$ have significant scatter, and it is difficult to differentiate between varying input models based on best-fit $R_{eff, meas}$ alone. We caution readers that fits to lower-SNR sources that are as compact as LRDs may not be reliable or physical.
    \item Two component decompositions performed by pysersic and GALFIT can only accurately measure the size of the underlying \sersic component when SNR is high and $R_{\textrm{eff,in}} \gtrsim1$. The two component fits can largely differentiate between compact and extended sources in estimating $f_{\textrm{PSF,fit}}$. This is an informative yet not perfect compactness criteria; at SNR$\sim50$ pixels, a pure point source becomes confused with a $R_{\textrm{eff,in}}\sim0.5$ pure-\sersic, while this happens when a \sersic only model has $R_{\textrm{eff,in}}\sim0.25$ pixels for images with $\text{SNR}\sim250$. Although $f_{\textrm{PSF,fit}}$ might not recover the exact model-$f_{\textrm{PSF,in}}$, it could be a good proxy for identifying extended emission in real LRDs.
    \item PSF fits show that nearly all of our simulated objects, regardless of intrinsic compactness, were indistinguishable from a point source in images with $\text{SNR}\lesssim25$. Even for higher-SNR images, only the most extended models showed significantly increased $\chi^{2}_{\nu}$ values when being fit by a PSF.
\end{itemize}

We also compared the pysersic parametric fits to the observed LRDs presented in \citet{koko24census} to those computed for our simulated LRD-like targets.  We present the best fit parameters for the \sersic only and two-component fits in Table \ref{tab:fit_results}. We found that almost all of the observed LRDs were best fit by \sersic profiles with $R_{\textrm{eff,fit}} \lesssim1$. For all of the observed LRDS, we also measured their $F_{0\farcs04}/F_{0\farcs02}$ ratios, or the compactness ratio used to initially select LRDs. We find that $R_{\textrm{eff,fit}}$ strongly correlates with compactness ratio across all SNR bins, for both the simulated and observed LRDs. A variety of simulated LRDs populate the same region of  $R_{\textrm{eff,fit}}$--$F_{0\farcs04}/F_{0\farcs02}$ parameter space as the observed LRDs. Matching each LRD to its ten nearest-neighbor models reveals that the scatter in model $R_{\textrm{eff,in}}$ is on the order of magnitude with the average, suggesting that there cold be a degree in morphological diversity in the observed LRD population. This is also consistent with what is revealed in the two component decomposition results; a vast majority of LRDs are extremely PSF-dominated ($f_{\textrm{PSF,fit}} \gtrsim 0.85$, but $15-20\%$ of the population have $f_{\textrm{PSF,fit}} \lesssim0.7$. 

Lastly, we also explored non-parametric compactness indicators. For each LRD, we compared its radial profile to the instrumental F444W PSF, as well as its best fit \sersic profile convolved with the PSF. We similarly found that $\sim 85\%$ or LRDs had F444W radial profiles that closely matched that of the PSF. We found that the remaining LRDs either were more extended than the PSF at all scales, or were extremely PSF-like towards the center of the profile but then exhibited significant extended emission at scales $R \gtrsim 5$ pixels, highlighting the underlying morphological diversity of of LRDs.

All of this taken together, it is likely that the total population of LRDs is not homogeneous. We stress that a vast majority of observed LRDs are almost completely unresolved with sizes $\lesssim50$ pc, which are consistent with the canonical picture of LRDs as being AGN-dominated. However, $\sim15\%$ of LRDs have evidence of some degree of extended F444W emission, either with $f_{\textrm{PSF}}\lesssim 0.7$ or measured sizes $\gtrsim 150 \text{pc}$. These same targets often have disturbed morphologies or multiple cores in the higher-resolution F115W images, suggesting that these could be the result of interacting pairs. Extremely compact, massive starburst galaxies with similar stellar densities exist at $\sim0.5$. These galaxies also host diffuse tidal features, indicative of having undergone a recent merger. It is possible that these galaxies are analogs to LRDS, or that the compact cores of LRDs are broad-line AGN embedded in the remnants of a galaxy interaction. Regardless, this subsample potentially opens the door to exploring the AGN-starburst connection at high-$z$.

\begin{acknowledgments}
K. E. W. and E. L.'s research was supported by an appointment to the NASA Postdoctoral Program at NASA Goddard Space Flight Center, administered by Oak Ridge Associated Universities under contract with NASA. We also thank Vasily Kokorev for useful conversation with regards to reproducing part of the \citet{koko24census} analysis as well as the COSMOS working group. 
\end{acknowledgments}

\bibliography{whalen_research}

\appendix
\section{Supplementary Figures}
\renewcommand{\thefigure}{\Alph{section}\arabic{figure}}
\counterwithin{figure}{section}

\begin{figure}[ht]
    \centering
\includegraphics[width=0.85\linewidth]{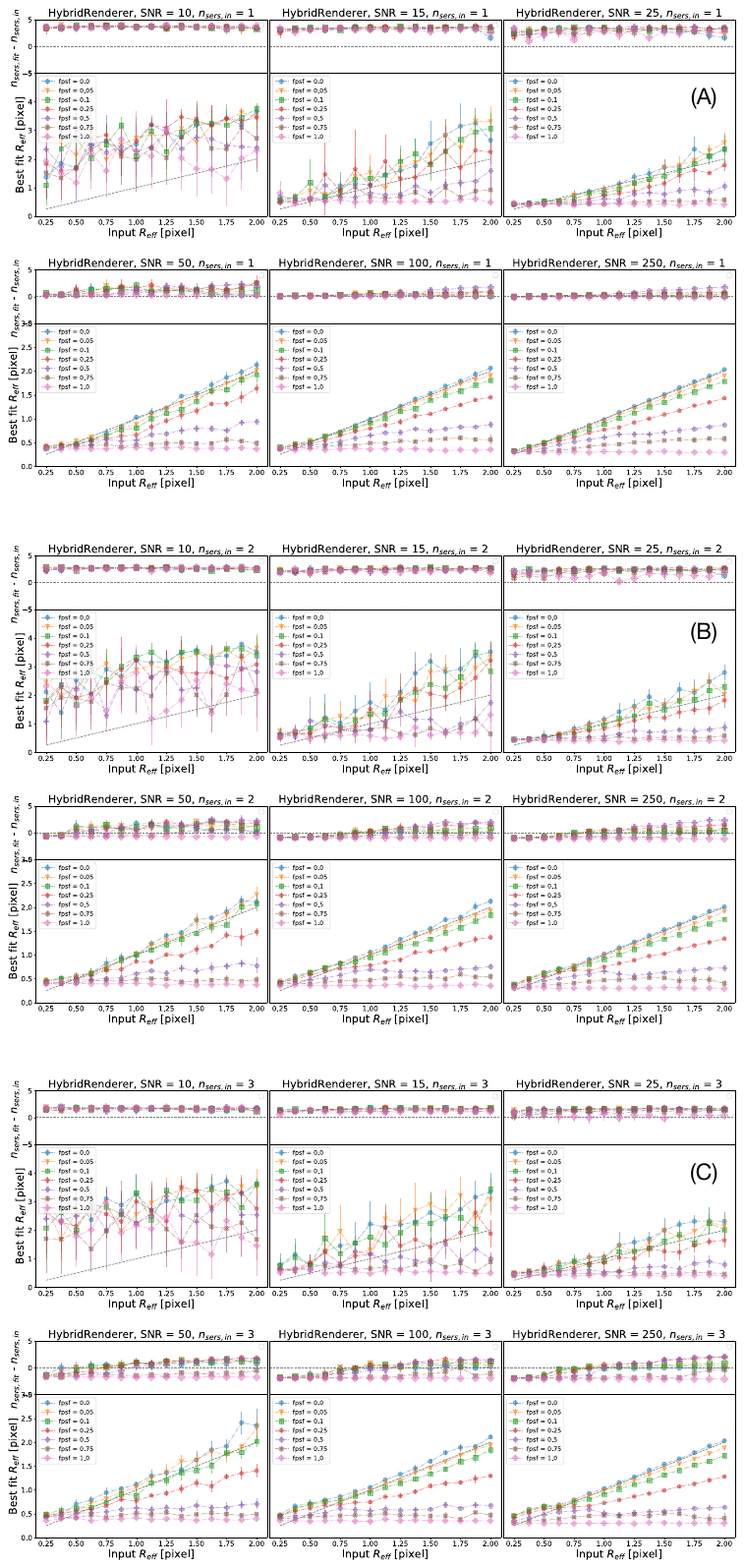}
    \caption{Best pysersic HybridRenderer\sersic fits to our suite of mock LRD-like objects, complementary to Figure \ref{fig:pysers_hyb_sers_n4}A. Each labeled panel here corresponds to different $n_{\textrm{sers,in}}$ values as given in the titles.}
    \label{fig:placeholder}
\end{figure}

\begin{figure}[htb]
    \centering
\includegraphics[width=\linewidth]{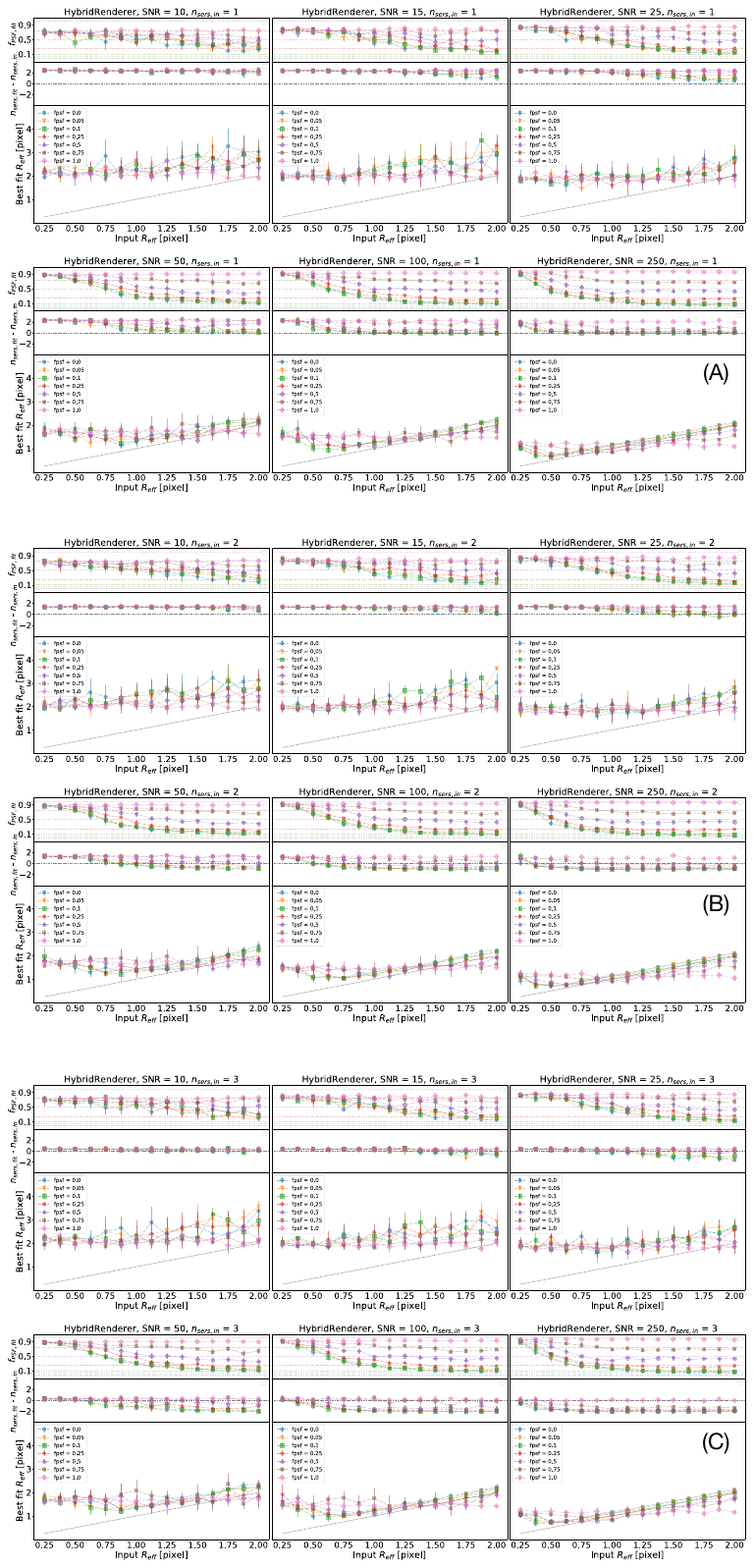}
    \caption{Best pysersic HybridRenderer \sersic + PSF fits to our suite of mock LRD-like objects, complementary to Figure \ref{fig:pysersic_hyb_ps+sers_n4}. Each labeled panel here corresponds to different $n_{\textrm{sers,in}}$ values as given in the titles.}
    \label{fig:placeholder}
\end{figure}

\begin{figure*}[htb]
    \centering
\includegraphics[width=\linewidth]{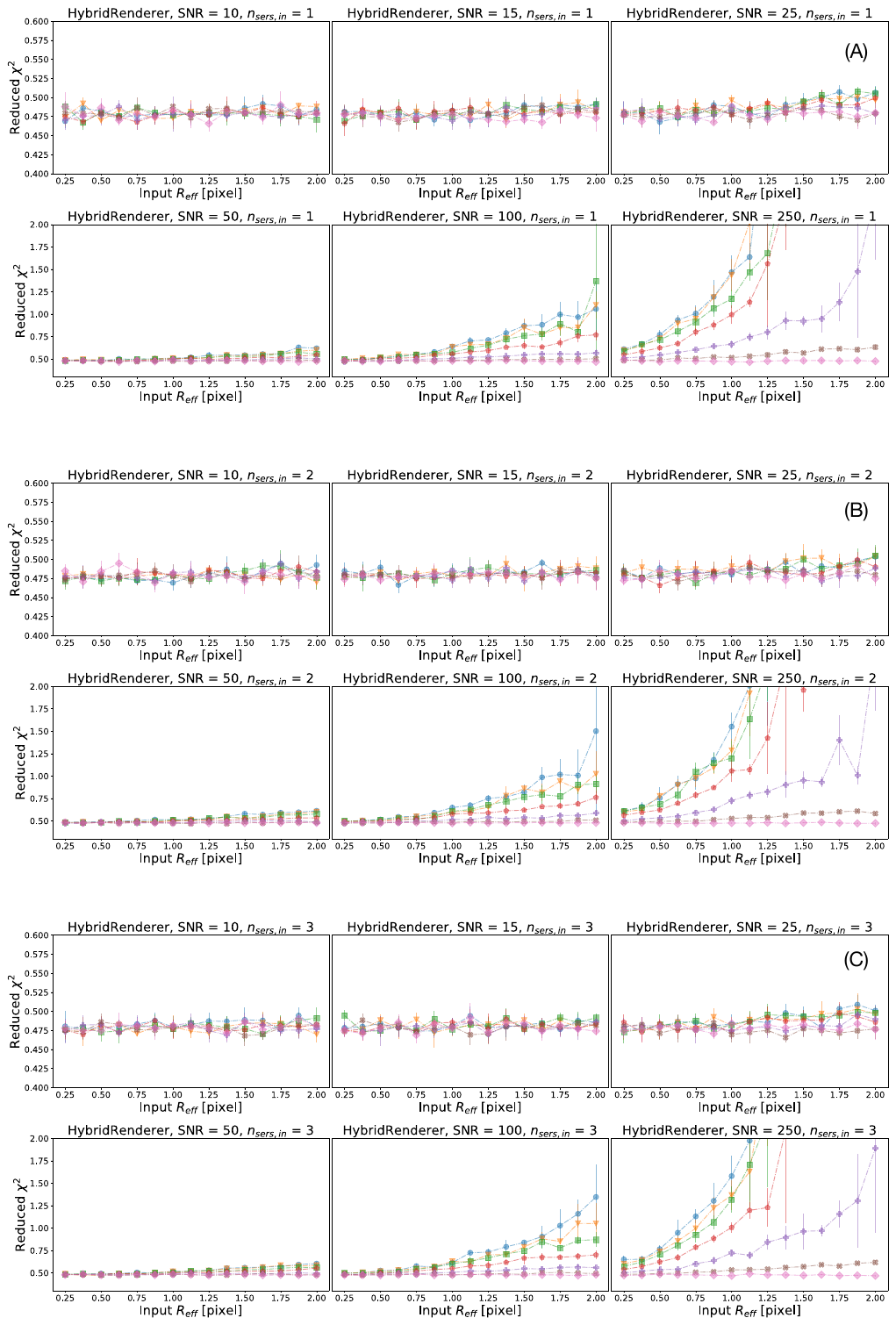}
    \caption{Best pysersic HybridRenderer PSF fits to our suite of mock LRD-like objects, complementary to Figure \ref{fig:pysersic_PSFfit}. Each labeled panel here corresponds to different $n_{\textrm{sers,in}}$ values as given in the titles.}
    \label{fig:placeholder}
\end{figure*}

\begin{sidewaysfigure*}[htb]
    \centering
\includegraphics[width=\linewidth]{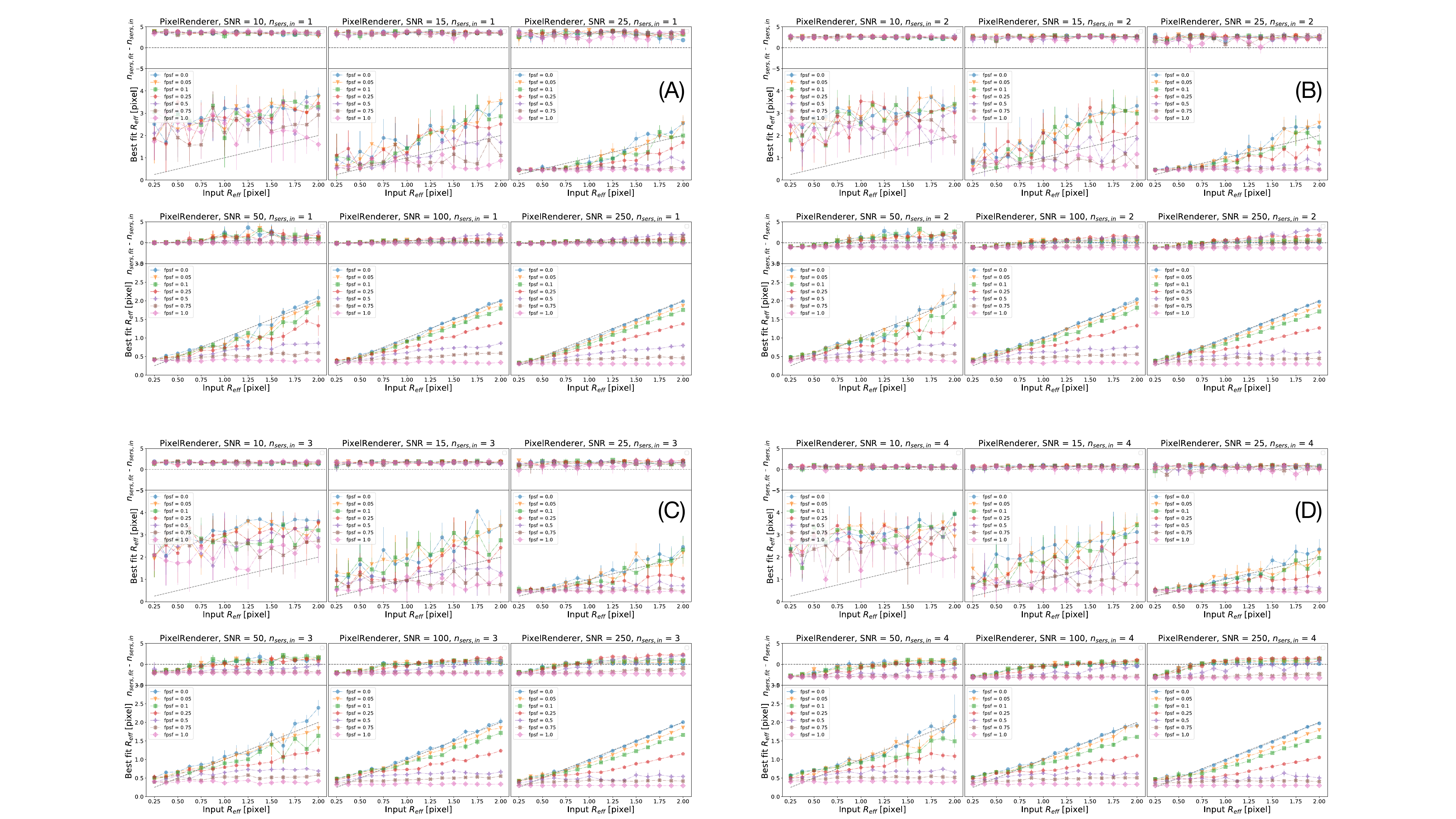}
    \caption{Best pysersic PixelRenderer \sersic fits to our suite of mock LRD-like objects, similar to Figure \ref{fig:pysers_hyb_sers_n4}A. Each labeled panel here corresponds to different $n_{\textrm{sers,in}}$ values as given in the titles.}
    \label{fig:placeholder}
\end{sidewaysfigure*}

\begin{sidewaysfigure*}[htb]
    \centering
\includegraphics[width=\linewidth]{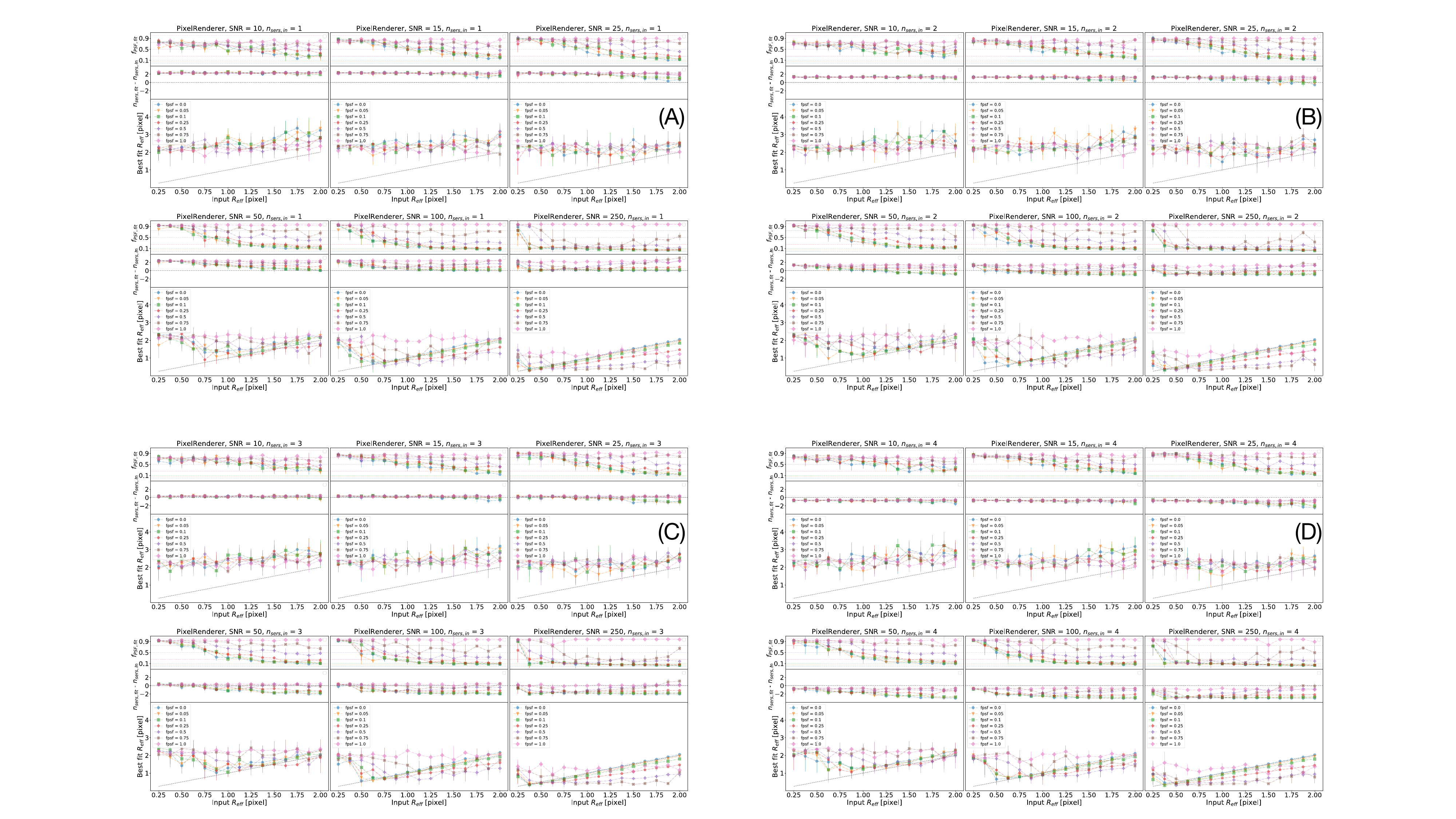}
    \caption{Best pysersic PixelRenderer \sersic+ PSF fits to our suite of mock LRD-like objects, complementary to Figure \ref{fig:pysersic_hyb_ps+sers_n4}. Each labeled panel here corresponds to different $n_{\textrm{sers,in}}$ values as given in the titles.}
    \label{fig:placeholder}
\end{sidewaysfigure*}

\begin{sidewaysfigure*}[htb]
    \centering
\includegraphics[width=\linewidth]{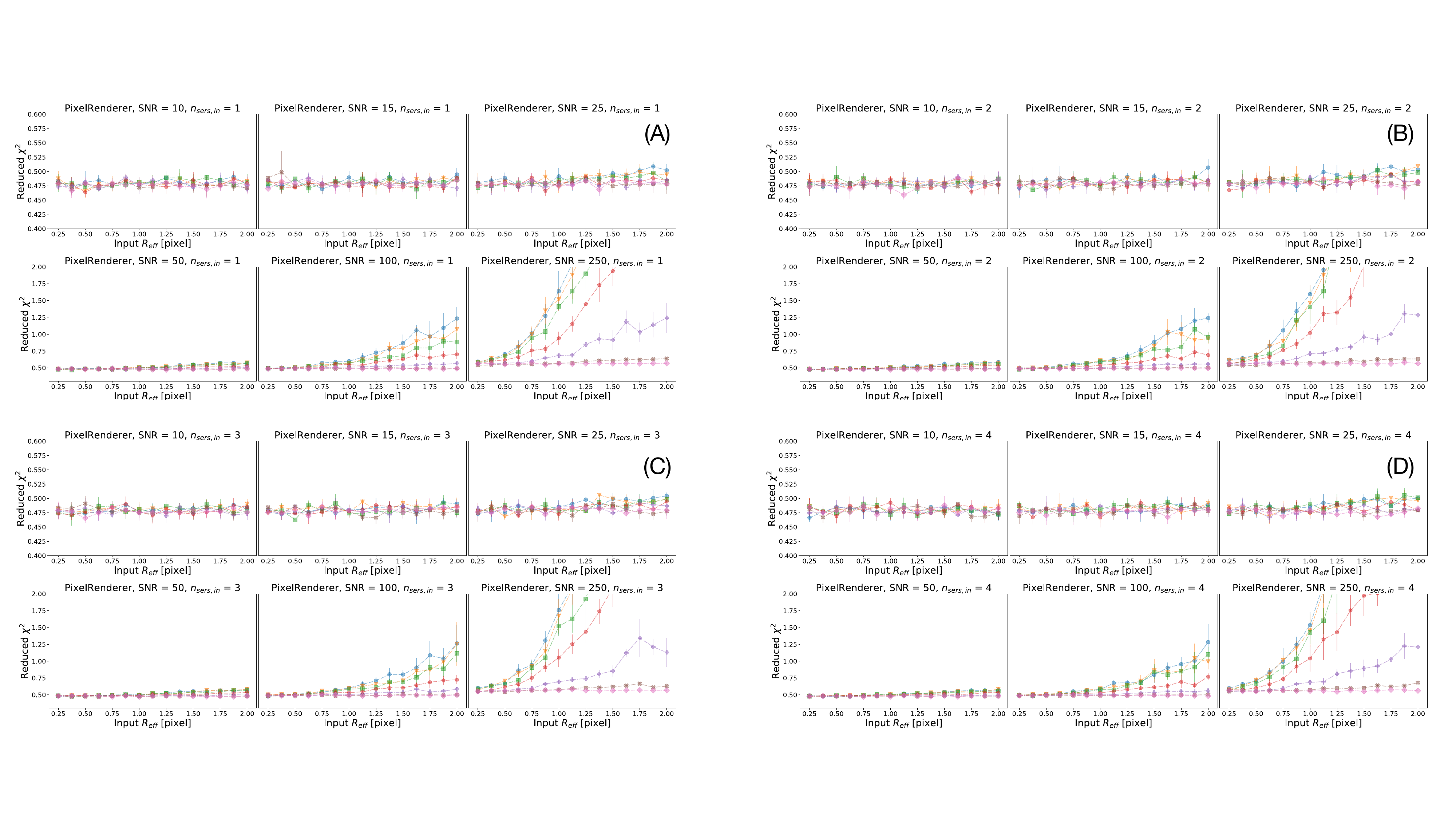}
    \caption{Best pysersic PixelRenderer PSF fits to our suite of mock LRD-like objects, similar to Figure \ref{fig:pysersic_PSFfit}. Each labeled panel here corresponds to different $n_{\textrm{sers,in}}$ values as given in the titles.}
    \label{fig:placeholder}
\end{sidewaysfigure*}

\begin{sidewaysfigure*}[htb]
\includegraphics[width=\textheight]{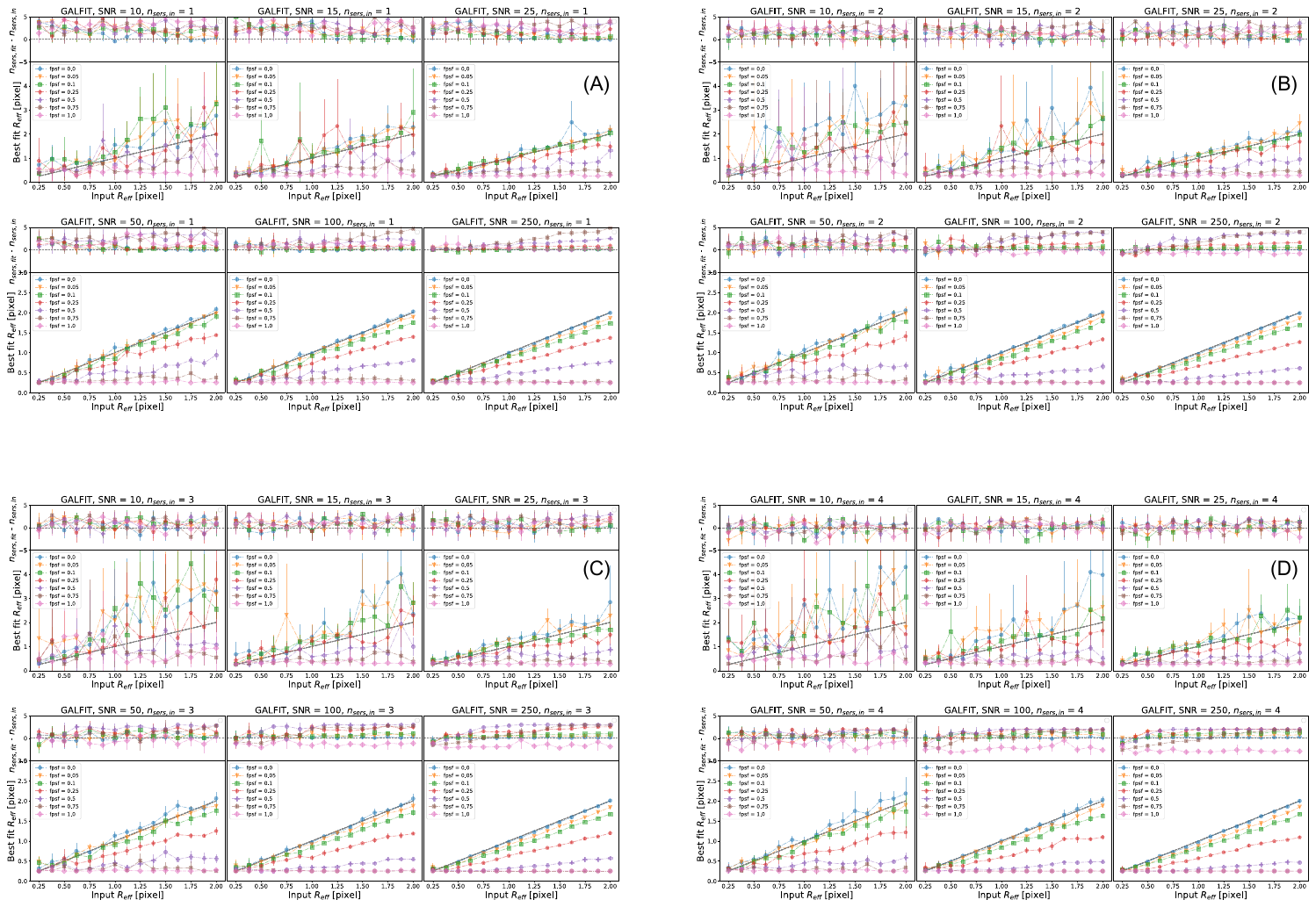}
\caption{Best GALFIT \sersic fits to our suite of mock LRD-like objects, similar to Figure \ref{fig:pysers_hyb_sers_n4}A. Each labeled panel here corresponds to different $n_{\textrm{sers,in}}$ values as given in the titles. These results are consistent with what we find for the pysersic HybridRenderer fits.}
\end{sidewaysfigure*}

\begin{sidewaysfigure*}[htb]
\includegraphics[width=\textheight]{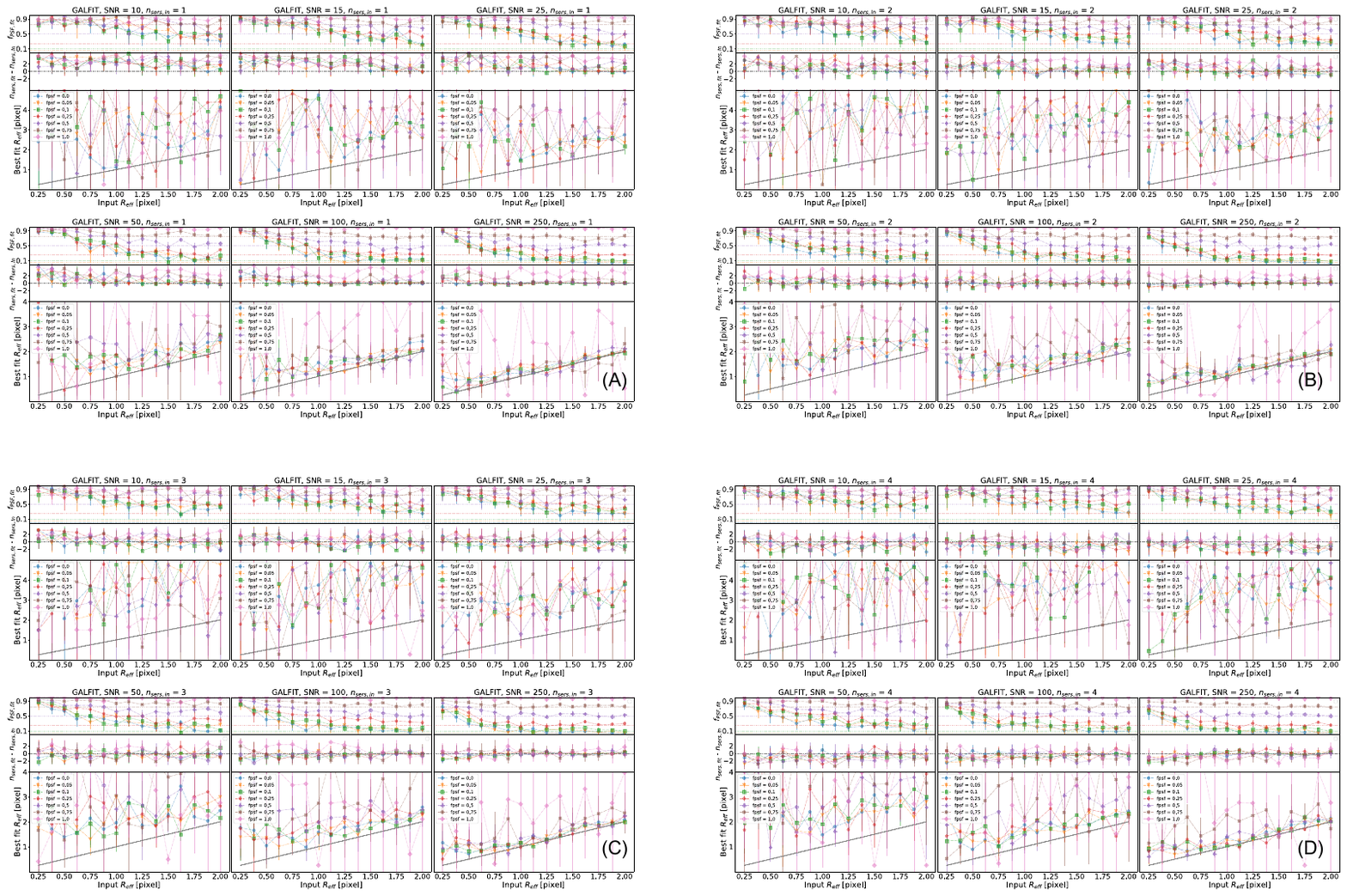}
\caption{Best GALFIT \sersic + PSF fits to our suite of mock LRD-like objects, similar to Figure \ref{fig:pysersic_hyb_ps+sers_n4}. Each labeled panel here corresponds to different $n_{\textrm{sers,in}}$ values as given in the titles. These results are consitent with what we find for the pysersic HybridRenderer fits.}
\end{sidewaysfigure*}

\begin{sidewaysfigure*}[htb]
\includegraphics[width=\textheight]{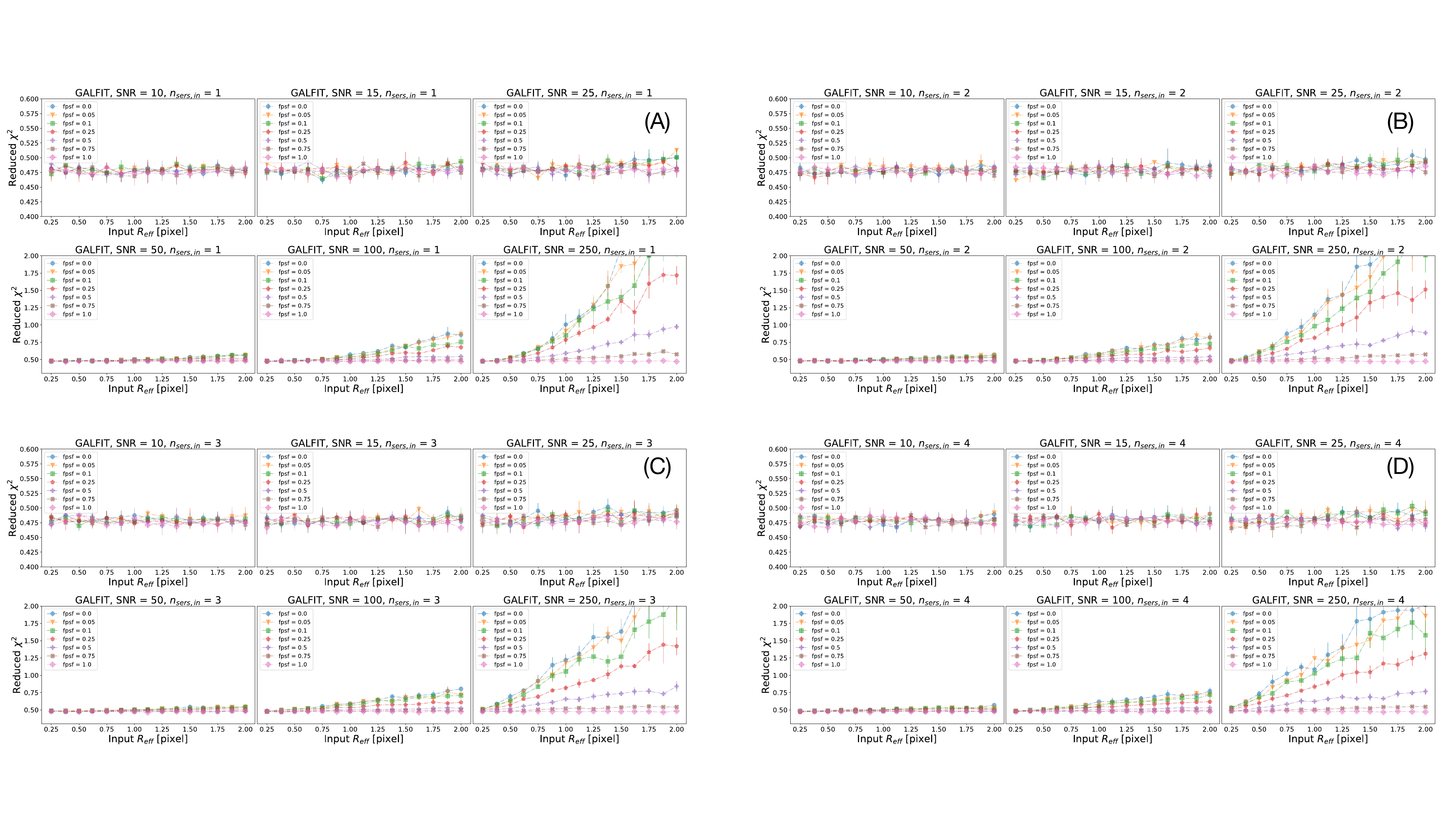}
\caption{Best GALFIT  PSF fits to our suite of mock LRD-like objects, similar to Figure \ref{fig:pysersic_PSFfit}. Each labeled panel here corresponds to different $n_{\textrm{sers,in}}$ values as given in the titles. These results are consitent with what we find for the pysersic HybridRenderer fits.}
\end{sidewaysfigure*}

\begin{figure}[ht]
    \centering
\includegraphics[width=1\linewidth]{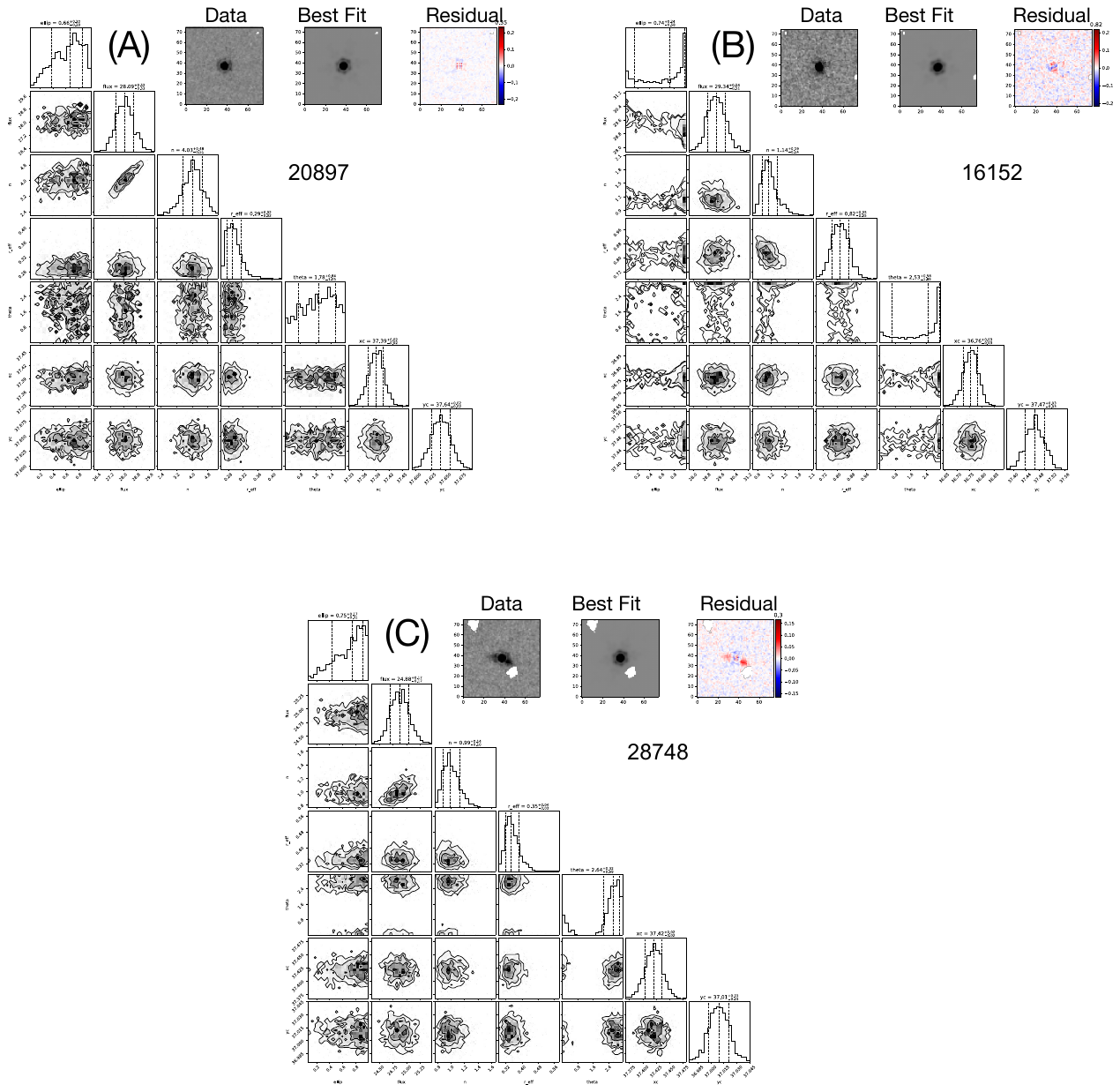}
    \caption{Best pysersic \sersic fits to LRDs 20897 \textit{(A)}, 16152 \textit{(B)}, and 28744 \textit{(C)}.}
    \label{fig:placeholder}
\end{figure}

\begin{figure}[ht]
    \centering
\includegraphics[width=0.9\linewidth,trim=2cm 4cm 2cm 1cm]{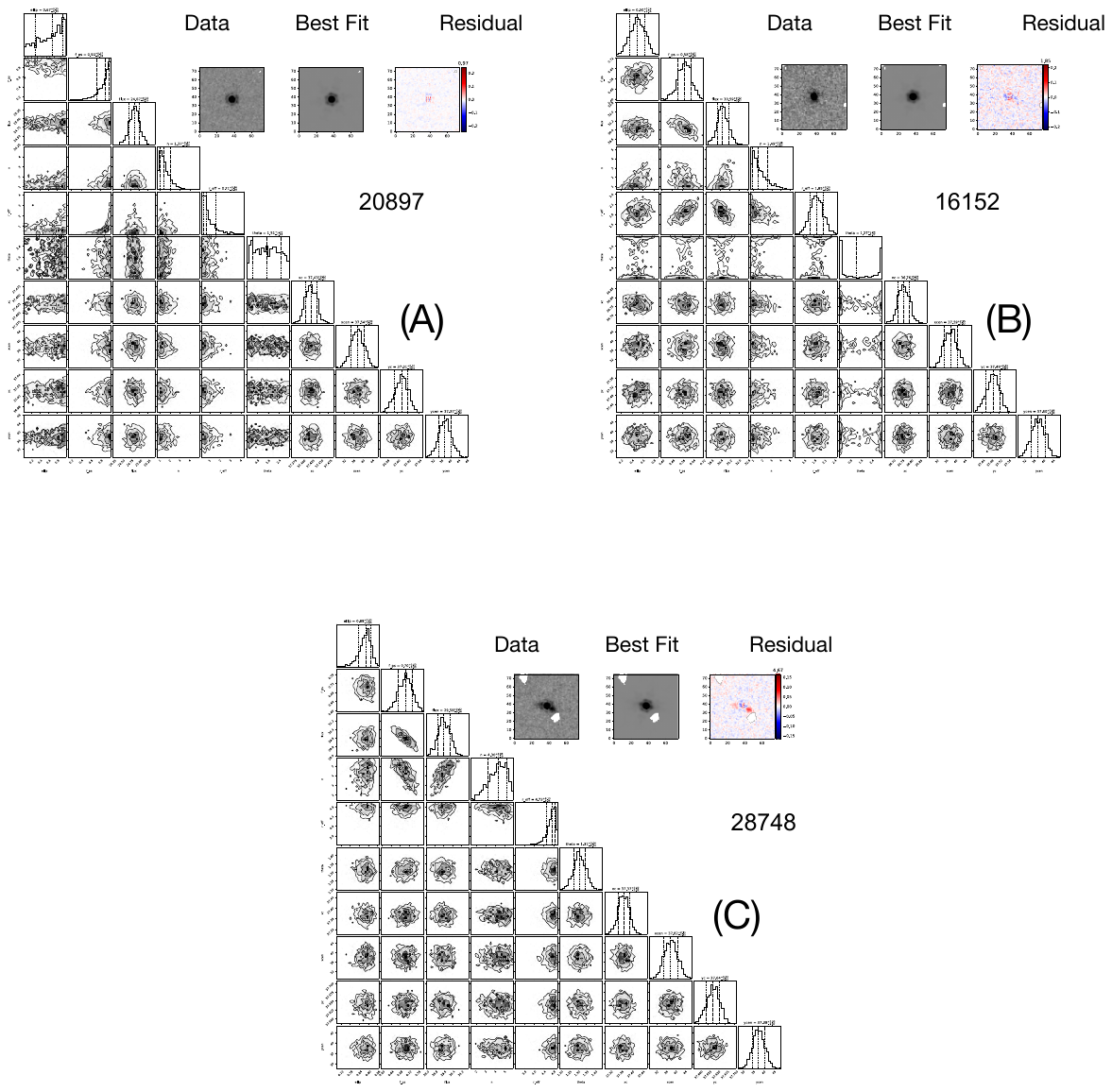}
    \caption{Best pysersic \sersic + PSF fits to LRDs 20897 \textit{(A)}, 16152 \textit{(B)}, and 28744 \textit{(C)}.}
    \label{fig:placeholder}
\end{figure}

\end{document}